\documentclass[prd,aps,nofootinbib,twocolumn]{revtex4-2}
\usepackage{amsmath,amssymb,amsfonts,color,graphicx,graphics,latexsym,placeins,epsfig,multirow}
\usepackage[toc]{appendix}
\usepackage{array}
\newcolumntype{P}[1]{>{\centering\arraybackslash}p{#1}}
\usepackage{footmisc}
\usepackage{mathrsfs}
\usepackage[compatibility=false]{caption}
\usepackage{subcaption}
\usepackage{comment}
\usepackage{bbold}
\usepackage{enumitem}
\setdescription{leftmargin=12.5pt}
\usepackage{bbold}
\usepackage{hyperref}
\usepackage{varioref}
\usepackage{color}
\DeclareMathOperator{\sech}{sech}

\usepackage{natbib}
\setcitestyle{square, comma, numbers, sort&compress}
\usepackage{float}
\newcommand*{\id}{{\rm\hbox{1\kern-0.15em \vrule width .1pt depth-.2pt}}}

\labelformat{section}{Section #1} 
\labelformat{subsection}{Section #1} 
\labelformat{subsubsection}{Section #1}
\labelformat{subsubsubsection}{Section #1}
\labelformat{equation}{Eq.~(#1)} 
\labelformat{figure}{Fig.~#1} 
\labelformat{subfigure}{Fig.~\thefigure#1} 
\labelformat{table}{Tab.~#1} 
\labelformat{appendix}{Appendix #1}
\begin{document}

\title{\Large \bf Gravitational wave memory in wormhole spacetimes}
\author{Indranil Chakraborty}
\affiliation{Department of Physics, Indian Institute of Technology Kharagpur, Kharagpur 721302, India}
\email{indradeb@iitkgp.ac.in}
\author{Soumya Bhattacharya}
\affiliation{Department of Astrophysics and High Energy Physics, S.N. Bose National Center for Basic Sciences, Kolkata 700106, India}
\email{soumya557@bose.res.in}
\author{Sumanta Chakraborty}
\affiliation{School of Physical Sciences, Indian Association for the Cultivation of Science, Kolkata 700032, India}
\email{tpsc@iacs.res.in}

\begin{abstract}

Gravitational wave (GW) memory is studied in the context of a certain class of braneworld wormholes. Unlike other wormhole geometries, this novel class of wormholes do not require any exotic matter fields for its traversability. First, we study geodesics in this wormhole spacetime, in the presence of a GW pulse. The resulting evolution of the geodesic separation shows the presence of displacement and velocity memory effects. Motivated by the same, we study the memory effects at null infinity using the Bondi-Sachs formalism, adapted for braneworld wormhole. Our analysis provides a non-trivial change of the Bondi mass after the passage of a burst of gravitational radiation and hence manifests the memory effect at null infinity. In both of these exercises, the presence of extra dimension and the wormhole nature of the spacetime geometry gets imprinted in the memory effect. Since future GW detectors will be able to probe the memory effect, the present work provides another avenue to search for compact objects other than black holes. 

\end{abstract}
\maketitle

\section{Introduction}

The direct detection of Gravitational Waves (GWs) from binary black hole and binary neutron star merger events \citep{Abbott:2016,Abbott:2017}, as well as the observations of the shadow of supermassive compact central objects e.g., the M87* and the SgrA* \citep{EHT:1,EHT:2,EHT:3,EHT:4,EHT:5,Sgr:1,Sgr:6}, are the two major observational breakthroughs in the field of gravitational physics within the last decade. Both of these observations depend crucially on the strong field behaviour of gravity and, in principle, can be used to test the robustness of General Relativity (GR) and also provide crucial pointers to the nature of the compact objects \citep{Yunes:2013,RAbbott_1:2021,Krishnendu:2021,Johannsen:2015,Ayzenberg:2018,Psaltis:2018,Banerjee:2019nnj, Chakraborty:2022zlq,Mishra:2019trb}. Despite the fact that --- so far GR has passed these strong field tests without any scar --- from a purely theoretical perspective, the inability of GR to correctly reconcile singularities occurring at the end point of gravitational collapse, or, as the starting point of our modern hot big bang cosmology, posits a serious challenge to the theory itself. This important shortcoming of GR must make room for investigating alternative possibilities, {\em vis-a-vis} modified near-horizon geometries. Among the alternatives, one can either look for higher curvature/higher dimensional theories of gravity, possibly emerging from some quantum gravity scenarios, or, modifications of the black hole paradigm itself. Such non-black hole compact objects can arise from quantum gravity-motivated theories e.g., fuzzballs \cite{Mathur:2005} or, compact objects made out of exotic matter fields, known as ECOs \cite{Cardoso:2016,Cardoso:2017,Cardoso:2019,Mazur:2001,Almheiri:2013}. Both of these classes of non-black hole objects, behave as a black hole for solar system tests of gravity, but appears to have distinct phenomenology, in contrast to that of the black hole, when they are probed using the strong field tests of gravity, in particular, GWs \cite{Lemos:2008,Pani:2008,Konoplya:2016,Rubio:2018,Abedi:2016,Mark:2017,Sennett:2017,Oshita:2018,Bueno:2017,Cardoso_ECO:2016,Dey:2020lhq,Dey:2020pth}.

{ However, the present generation of GW detectors are not sensitive enough to detect the modifications in the GW waveform originating from such non-black hole objects. In absence of such definitive tests that can either confirm or nullify the existence of these non-black hole objects, it is necessary to study the strong-gravity signatures of these exotic objects in order to gain a better understanding of their physical characteristics. Here we attempt to study some properties associated with GWs in the background of one of the oldest as well as interesting class of ECOs, {\em viz.} wormholes.} These are spacetimes joining two distinct universes by a throat \cite{Morris:1988,Bronnikov:2002,Bronnikov:2003,Tsukamoto:2012,Ohgami:2015,Shaikh:2018,Banerjee:2019,DuttaRoy:2019,Bronnikov:2020,Franzin:2022}. If one can travel from one universe to the other, it will be referred to as a traversable wormhole and this, in general, requires exotic matter fields. However, there exist one class of wormholes, which do not require exotic matter for its existence, known as the braneworld wormholes \cite{Kar:2015}, corresponding to Randall-Sundrum two braneworld scenario \cite{RS1}. This has two advantages --- (a) the presence of higher dimension is imprinted on this wormhole solution and hence can possibly be tested using GWs or, black hole shadow measurements, (b) the wormhole does not require any exotic matter on the four-dimensional brane, the contributions from higher dimensions take care of the exotic nature to support traversability of the wormhole. Therefore, it is worthwhile to study various properties of this wormhole and hence look for its observational signatures. This is because, it will not only validate the existence of wormholes, but also of compact extra dimensions, thus providing a unified testing ground for non-black hole nature of compact objects, {\em vis-a-vis} of theories beyond GR. Several aspects of this braneworld wormhole, e.g., its stability under perturbations due to various fields, in particular the ringdown structure \cite{Biswas:2022}, as well as implications for black hole shadow measurements \cite{Banerjee:2020}, have already been explored. Intriguingly, all of these explorations have provided quite promising results and in particular, demonstrates the existence of echoes in the GW signal. In this work, we wish to explore another direction, which has remained largely unexplored, but holds immense potential for the future generations of GW detectors, namely the gravitational memory effect.  

With the improvement in the detection prospects for the future ground based GW detectors and also the launch of the space based-detector LISA in the near future, may provide us an opportunity to observe {\em GW memory effect} \cite{Boersma:2020}. The memory effect brings in both the strong-field, as well as non-linear aspects of general relativity, which is yet to be observed. It refers to the lasting change in the relative distance between test particles when a GW passes through the ambient spacetime \cite{Braginsky:1985,Favata:2010}. Being a subtle DC shift to the overall GW amplitude, it has remained undetected in the past observations taken by LIGO \cite{Hubner:2021}. There have been proposals of stacking the known GW signals observed by LIGO-Virgo in order to detect this effect \cite{Lasky:2016}. Initially studied in the context of hyperbolic scattering \cite{Zeldovich:1974} and gravitational bremsstrahlung \cite{Kovacs:1978}, the memory effect was also sho \cite{Christodoulou:1991}. Recent works on memory effects involve, generalization to electrodynamics \cite{Bieri:2013,Winicour:2014} and Yang-Mills theories \cite{Pate:2017,Jokela:2019}, investigating features of extra dimensions \cite{Hollands:2017,Wald:2018,Ferko:2021}, distinguishing modified theories of gravity from GR, e.g., scalar-tensor theories \cite{Du:2016,Hou:2020JHEP,Hou:2020,Tahura:2020} and Chern-Simons gravity \cite{Hou:2022}. Moreover, there have also been works generalizing memory effects to the symmetries associated with near horizon geometries for black holes \cite{Donnay:2018, Srijit:2019,Wald:2020}. In this work, we wish to study the gravitational memory effect for braneworld wormholes, a class of ECOs, for the first time. Our aim, in the present context, is to infer the imprints of the non-black hole nature of this static wormhole spacetime and the presence of extra dimensions on the memory effect. We attempt this exercise of finding the memory effects in two distinct ways --- (a) we perform a geodesic analysis (as worked out in \cite{Braginsky:1985,Zhang:2017soft}) and comment on the presence of displacement and velocity memory effect in the background wormhole spacetime by introducing an additional GW perturbation, (b) we study memory effects at null infinity using the Bondi-Sachs formalism \citep{Madler:2016}. 

The organization of the paper is as follows: in \ref{review_braneworld} we briefly review the wormhole geometry in the context of braneworld scenario. \ref{memory_geodesic} deals with the study of geodesics and subsequent analysis of displacement and velocity memory effects and finally in \ref{null_memory}, we will study the influence of a perturbing GW on the wormhole metric in the Bondi-Sachs gauge and the associated null memory effects. We conclude with a discussion on the results obtained and provide an outline of the future directions to pursue. \\

\section{Brief review of wormhole geometry in the braneworld scenario}\label{review_braneworld}

Let us now discuss briefly the braneworld model under investigation. In the model, we have a five dimensional spacetime (referred to as the bulk), within which two four dimensional branes are embedded. The extra dimension is spacelike in nature and is described by the coordinate $y$. One of the brane is located at $y=0$, and dubbed as the Planck brane, while the other one is located at $y=\ell$, referred to as the visible brane. The proper distance between the two branes is given by the integral of the $g_{yy}$ component over the extent of the extra dimension, yielding $d(x)=e^{\phi(x)}\ell$, where the field $\phi(x)$ is referred to as the radion field. Since we are interested in the measurements done by a four-dimensional observer, it will suffice to consider the low energy effective gravitational field equations. This can be achieved by projecting the five dimensional Einstein's equations on the four dimensional brane, and then expanding the same in the ratio of the (bulk/brane) curvature length scale. Thus we finally obtain the following gravitational field equations on the visible brane \cite{Kanno:2002iaa,Kar:2015},
\begin{widetext}
\begin{equation}\label{eq:field_equation}
\begin{split}
G_{\mu \nu}&=\frac{\kappa_{5}^2}{\ell \Phi} T^{\rm V}_{\mu \nu} + \frac{\kappa_{5}^2 (1+ \Phi)}{\ell \Phi} T^{\rm P}_{\mu \nu}+\frac{1}{\Phi} \Big(\nabla_{\mu} \nabla_{\nu} \Phi - g_{\mu \nu} \nabla^{\alpha} \nabla_{\alpha} \Phi  \Big)-\frac{3}{2 \Phi (1+ \Phi)} \Big( \nabla_{\mu} \Phi \nabla_{\nu} \Phi - \frac{1}{2} g_{\mu \nu} \nabla^{\alpha} \nabla_{\alpha} \Phi \Big)~. 
\end{split}
\end{equation}
\end{widetext}
Here $g_{\mu \nu}$ is the visible brane metric, $\nabla_{\mu}$ is the covariant derivative with respect to $g_{\mu \nu}$ and $\kappa_{5}^2$ is the five dimensional gravitational coupling constant. Moreover, $T^{\rm P}_{\mu \nu}$ and $T^{\rm V}_{\mu \nu}$ are the energy momentum tensors on the Planck brane and the visible brane, respectively. The scalar field $\Phi$, appearing in \ref{eq:field_equation} is defined as, $\Phi\equiv \exp[2e^{\phi(x)}]-1$, where $\phi(x)$ is the radion field. { The energy density of the on-brane matter field must be small compared to the brane tension in order for the low-energy effective theory to hold. Wormholes, being non-singular solutions, have finite energy density and pressure for the on-brane matter fields everywhere, thereby ensuring the validity of this theory in the contexts of wormhole geometry \cite{Kar:2015}.}

{ In addition, it must be noted that the wormholes require exotic matter fields for their traversability, at least in the context of GR. This is because the violation in the convergence condition for timelike geodesics, leads to a violation of the energy conditions for the stress-energy tensor sourcing the wormhole geometry \cite{Morris:1988}. However, in the case of braneworld scenario, the total energy-momentum tensor has contributions from the matter present on the two 3-branes (visible and planck) and a geometric stress due to the radion field generated from the bulk spacetime. Hence, one can sustain this wormhole geometry with the on-brane matter satisfying the energy conditions and the violations of the energy conditions for the total energy-momentum tensor can be attributed to the bulk spacetime (similar situations may arise in the context of scalar coupled Gauss-Bonnet gravity, see e.g., \cite{Antoniou:2019awm}). Thus the resulting wormhole solution will be constructed out of normal matter fields on the brane, with exotic matter field on the bulk. Since we cannot access the energy scale of the bulk spacetime, the existence of such exotic matter in the bulk is not of much concern to our present analysis. Moreover, such braneworld wormholes have also been shown to be stable under scalar, electromagnetic and axial gravitational perturbations in \cite{Biswas:2022}.}

In order to avoid non-locality in the theory, i.e., we do not want the dynamics of the visible brane to be governed by the energy momentum tensor of the Planck brane, we will work with $T^{\rm P}_{\mu \nu}=0$ i.e., there is no matter on the Planck brane. With this choice, the field equation for $\Phi$ takes the following form,
\begin{equation}
\nabla^{\alpha} \nabla_{\alpha} \Phi = \frac{\tilde p^2}{\ell} \frac{T^{\rm V}}{2 \omega + 3} - \frac{1}{2 \omega + 3} \frac{d\omega}{d\Phi}\Big(\nabla^\alpha \Phi \Big)\Big(\nabla_\alpha \Phi \Big)~,
\end{equation}
where, $T^{\rm V}$ is the trace of the energy momentum tensor on the visible brane and the coupling function $\omega(\Phi)$, is defined as,
\begin{equation}
\omega(\Phi)=-\frac{3 \Phi}{2(1 + \Phi)}~.
\end{equation}
Thus, the low energy effective braneworld scenario can be written as a generalized Brans-Dicke (BD) \cite{Faraoni:2004} theory, with a variable BD parameter $\omega(\Phi)$. Given these, we rewrite the gravitational field equations as,
\begin{equation}
G_{\mu\nu}=\frac{\kappa_{5}^2}{l\Phi} T_{\mu\nu}^V+\frac{1}{\Phi}T_{\mu\nu}^\Phi~. \label{eq:einstein_tensor}
\end{equation}
Here, $T_{\mu\nu}^\Phi$ is to be identified with the sum of the third and the fourth terms of the right hand side of \ref{eq:field_equation}, without the $(1/\Phi)$ part.

The above set of equations for the metric functions, as well as for the scalar field $\Phi$, can be solved assuming a static and spherically symmetric metric ansatz along with an anisotropic fluid source on the visible brane with vanishing trace. This simplifies the problem of solving the field equations a lot and one arrives at a two-parameter family of solutions with $R=0$, written in the Schwarzschild-like coordinates as \cite{SK:2002,Kar:2015},
\begin{align}
ds^{2}&=-\left(\kappa +\lambda \sqrt{1-\frac{2M}{r}}\right)^2 dt^{2}
+\left(1-\frac{2M}{r}\right)^{-1}dr^2 
\nonumber
\\
&\hskip 1 cm +r^2 \Big(d\theta^2 + \sin \theta^2 d\phi^2 \Big)~.
\label{le1}
\end{align}
{ Here, in the limit, $\kappa=0$ and $\lambda=1$, we get back the Schwarzschild geometry. Note that in \ref{le1}, the $tt$-component of the metric is not given in the standard asymptotic form, since it does not reduce to unity in the limit of $r\rightarrow \infty$, rather to $(\kappa+\lambda)^{2}$. Therefore, we rescale the time coordinate by $t \to t/(\kappa+\lambda)$ and define the ratio $\frac{\kappa}{\lambda}=p$, since this is the only parameter which characterises the wormhole geometry. Finally, using the coordinate transformation $u=t-r_{*}$, where $r_{*}$ is the Tortoise coordinate, defined as $(dr_{*}/dr)=(1/\sqrt{-g_{tt}g^{rr}})$, the metric in \ref{le1} becomes,
\begin{equation}
ds^2=-f(r) ~ du^2 -2g(r) ~ dudr +r^2 ~ d\theta^2 + r^2 ~\sin^2\theta\, d\phi^2~. 
\label{wormhole_line_element}
\end{equation}
Here we have denoted the $uu$ and $ur$ components of the metric as $f(r)$ and $g(r)$, with the following expressions for them,
\begin{align}
f(r)&\equiv \bigg(\frac{p}{p+1}+\frac{1}{1+p}\sqrt{1-\frac{2M}{r}}\bigg)^2~, 
\label{fr}
\\
g(r) &\equiv \bigg(\frac{p}{p+1}+\frac{1}{1+p}\sqrt{1-\frac{2M}{r}}\bigg)\bigg(1-\frac{2M}{r}\bigg)^{-1/2}~.
\label{gr}
\end{align}
The scalar field, on the other hand, is best written in the isotropic coordinate $r'$, such that
\begin{equation}
\Phi(r')=\Big(\frac{C_1}{M} \log \frac{2r'q+M}{2r'+M}+C_4\Big)^2 -1 ~,
\label{eq:scalar_field_isotropic}
\end{equation} 
with the isotropic coordinate $r'$ being related to the Schwarzschild coordinate $r$ through the following relation: $r=r'(1+\frac{M}{2r'})^{2}$.} Note that the two coordinates $r$ and $r'$ become identical in the asymptotic limit. Moreover, $C_1$ and $C_4$, appearing in \ref{eq:scalar_field_isotropic}, are positive non-zero constants and $q=\{(p+1)/(p-1)\}$, where $p$ is the wormhole parameter, defined earlier. 

Unlike the Schwarzschild spacetime, here the radial coordinate $r$ can be extended only upto $r=2M$ and it is also evident from \ref{le1} that as long as $\kappa$ is non-zero and positive, $g_{tt} \neq 0$ for all  $r\geq2M$. This suggests that there is no event horizon in this spacetime. Though, the surface $r=2M$ is not an event horizon, it is indeed null, as $g^{rr}$ vanishes there and hence in the above solution as well $r=2M$ is a special surface, and is referred to as the throat of the wormhole. Physically, the above solution depicts two separate universes connected together at the throat, located at $r=2M$, which is traversable. The expression for the anisotropic fluid matter at the throat, necessary for traversability can be obtained from \cite{Kar:2015}. We do not provide it here, since this is not required for the computation of gravitational memory, which is the prime goal in the present context. The above provides a broad introduction to the wormhole geometry we will be working with and we shall apply these results in order to compute the memory effect, to be discussed below.


\section{Memory of geodesics}\label{memory_geodesic}

{ In this section, we will present the analysis of the memory effect vis-a-vis the geodesic deviation between neighbouring geodesics due to a propagating GW, with the geodesic separation quantifying the amount of displacement memory effect. Moreover, if the geodesics do not have constant separation, after the passage of the GW pulse, one can also associate a velocity memory effect with these geodesics as well. }

Such effects have been studied in the recent past \cite{Zhang:2017,Zhang:2017soft,Flanagan:2019,Chak:2020,Chak1:2020} by investigating the evolution of geodesics in exact plane GW spacetimes. By choosing a Gaussian pulse for the polarization (radiative) term in the line element of the plane GW spacetime, the geodesic equations can be solved numerically, and then the change in the separation and velocity (the displacement and the velocity memory, respectively) can be computed due to the passage of the GW pulse. Lately, the above formalism has also been generalized in the context of alternative theories of gravity \citep{Siddhant:2020,Chak:2022} to look for signatures of such alternative theories in the displacement and velocity memory effects. In the present work, we will study the evolution of the geodesics in the wormhole background presented above, in the presence of a GW pulse and study how the displacement and velocity memory effects depend on the wormhole nature of the background geometry. { For this purpose, we write down the spacetime metric as a sum of the background wormhole geometry $g_{\mu \nu}$ and a GW perturbation $h_{\mu \nu}$, such that the line element becomes,
\begin{equation}
ds^2= (g_{\mu\nu} + h_{\mu\nu})\, dx^\mu \, dx^\nu~. 
\end{equation}
As we have already mentioned, $g_{\mu\nu}$ is the wormhole metric given in \ref{wormhole_line_element} and $h_{\mu\nu}$ is the GW perturbation. Using the wormhole geometry presented in \ref{wormhole_line_element} explicitly, The resulting geometry becomes, 
\begin{align}
ds^2&=-f(r)\, du^2 -2g(r)\, du dr +\left[r^2 + r H(u)\right]~ d\theta^2 \nonumber
\\
&\hskip 0.5 cm + \left[r^2 - r H(u)\right]~\sin^2\theta\, d\phi^2~,
\end{align}
where, $f(r)$ and $g(r)$ are given by \ref{fr} and \ref{gr}, respectively. The function $H(u)$ corresponds to the GW pulse and we have assumed that $h_{\mu \nu}$ can be expressed in the transverse-traceless (TT) gauge. In order to generate this perturbation in the wormhole metric, we need to include in the matter sector an energy momentum tensor, which can source the GW pulse $H(u)$ in the TT gauge. Such an energy momentum tensor can arise from an expanding anisotropic fluid shell. Prior to the perturbation, the fluid was non-dynamical and the $u$-constant hypersurfaces were spherically symmetric. Due to this expansion, the GW pulse is generated and it propagates over the wormhole spacetime to future null infinity. In what follows, we will derive the displacement and the  velocity memory effects as the propagating GW crosses a congruence of timelike geodesics in the wormhole background, leading to a change in the separation between these comoving geodesics.}

The GW pulse profile described by $H(u)$ is taken to be, $H(u) = A\sech^2 (u-u_0)$, where $A$ denotes the amplitude of the GW pulse and it is centered around $u=u_0$. Since the above wormhole spacetime along with the GW pulse respects the spherical symmetry, we can choose the equatorial plane, located at $\theta=(\pi/2)$, to be the plane on which all the geodesics are located. Therefore, on the equatorial plane, the geodesic equations for the $u$ coordinate becomes,
\begin{align}
\ddot u &- \frac{f'}{2g}  \dot u^2 - \frac{H - 2r}{2g} \dot \phi^2 = 0~,
\label{udot}
\end{align}
Along identical lines one can arrive at the geodesic equations for the other coordinates on the equatorial plane, in particular, the respective geodesic equations for the $r$ and the $\phi$ coordinates are given as, 
\begin{align}
\ddot r &+ \frac{f}{2g^2} f' \dot u^2 + \frac{f'}{g} \dot r \dot u + \frac{g'}{g}\dot r^2 
\nonumber
\\
&\hskip 2 cm + \frac{fH - 2fr -rg H'}{2g^2} \dot \phi^2 =0 ~,
\label{rdot}
\\
\ddot \phi &- \frac{H'}{r-H} \dot \phi \dot u + \frac{2r - H}{r(r-H)} \dot \phi \dot r = 0~.
\label{fdot}
\end{align}
As mentioned before, all the above equations are written on the $\theta = \frac{\pi}{2}$ plane and here `overdot' denotes derivative with respect to the proper time $\tau$ associated with the geodesics, while `prime' denotes derivative with respect to the argument of the respective function. For example, $H'\equiv (dH/du)$ and $f'=(df/dr)$. 
\begin{figure*}[t]
	\centering
	\begin{subfigure}[t]{0.45\textwidth}
		\centering
		\includegraphics[width=\textwidth]{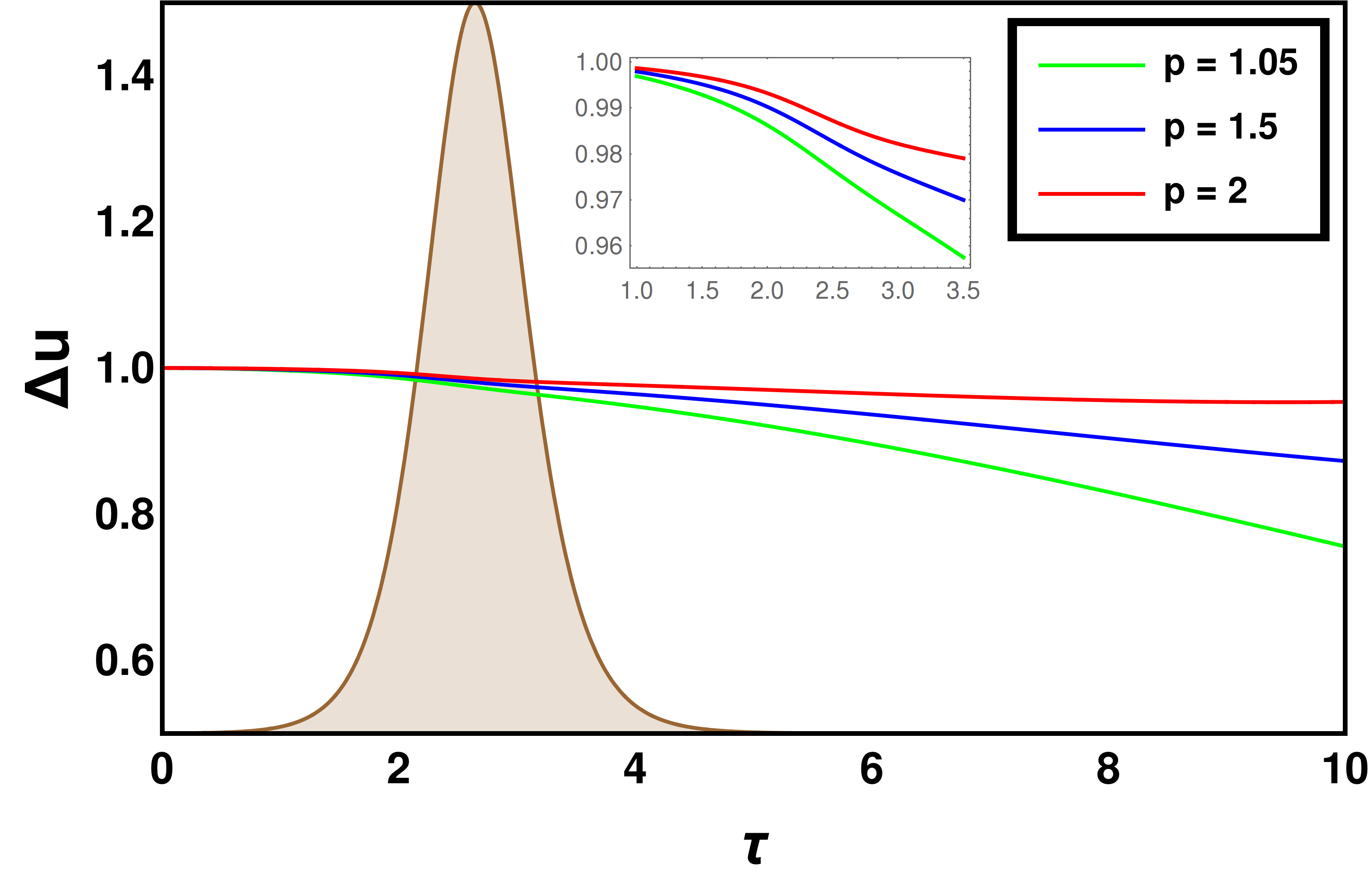}
		 \caption{\centering{\small Variation of the difference between null coordinate of the two timelike geodesics, denoted by $\Delta u$ with respect to the proper time $\tau$, for different values of the higher dimensional parameter $p$ has been presented.}}
		\label{fig:kappau}
\end{subfigure} \hspace{1cm}
	\begin{subfigure}[t]{0.45\textwidth}
		\centering
		\includegraphics[width=\textwidth]{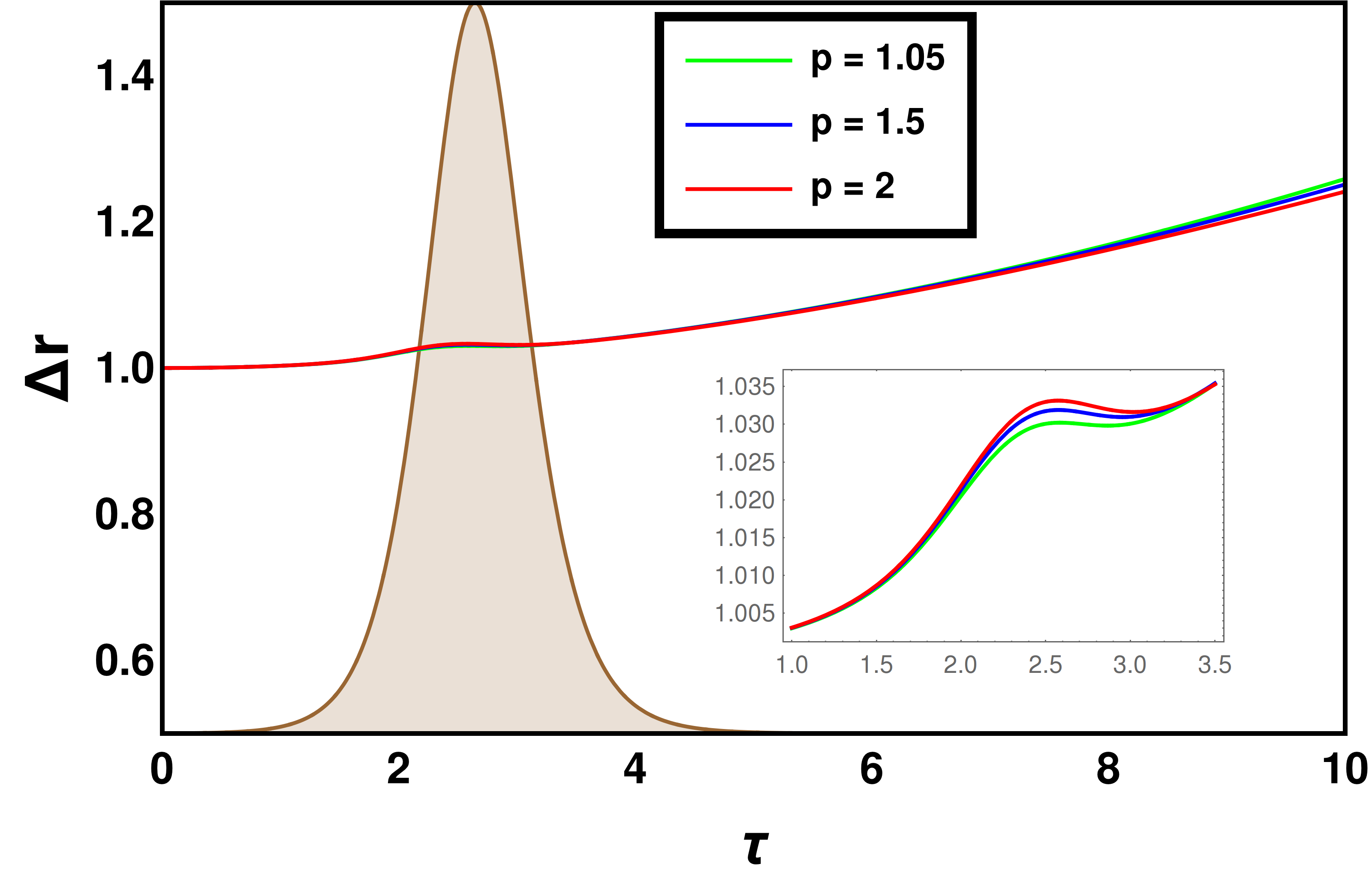}
		 \caption{\centering{\small Variation of the difference between the radial coordinates of the timelike geodesics, namely $\Delta r$ has been presented against the proper time $\tau$ for different values of the wormhole parameter $p$.}}
		\label{fig:kappar}
	\end{subfigure}
	\caption{\centering{{\small Variation of the difference of both the coordinates between the timelike geodesics, namely $\Delta u$ and $\Delta r$ has been plotted with the proper time for different choices of the wormhole parameters.}}}
	\label{fig:Variation_du_dr}
\end{figure*}
\begin{figure*}[t]
	\centering
	\begin{subfigure}[t]{0.45\textwidth}
		\centering
		\includegraphics[width=\textwidth]{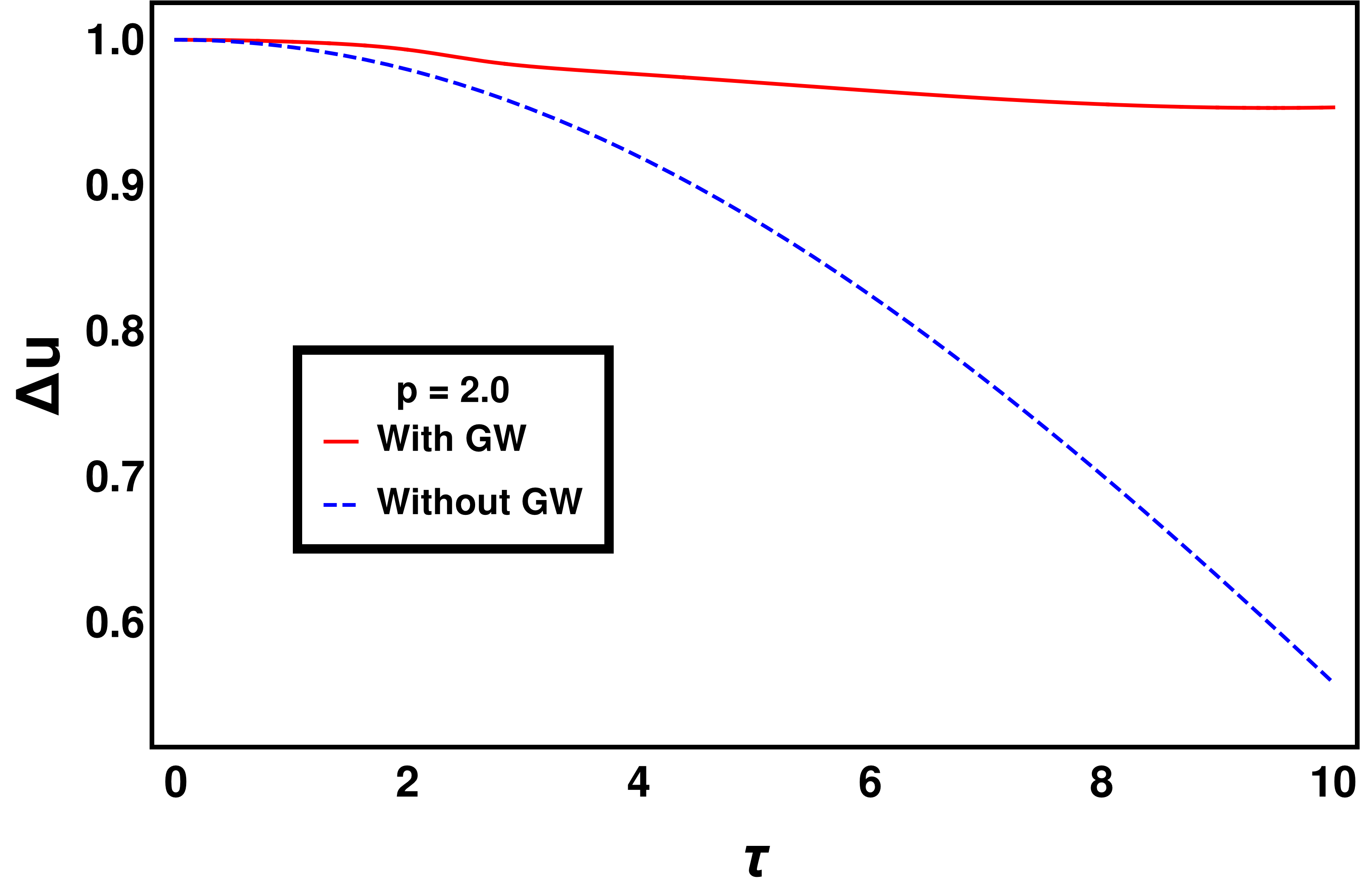}
		 \caption{\centering{\small Variation of the difference between null coordinate of the two timelike geodesics, denoted by $\Delta u$, has been presented against the proper time $\tau$, in the presence of and in the absence of the GW pulse.}}
		\label{fig:uwgw}
\end{subfigure} \hspace{1cm}
	\begin{subfigure}[t]{0.45\textwidth}
		\centering
		\includegraphics[width=\textwidth]{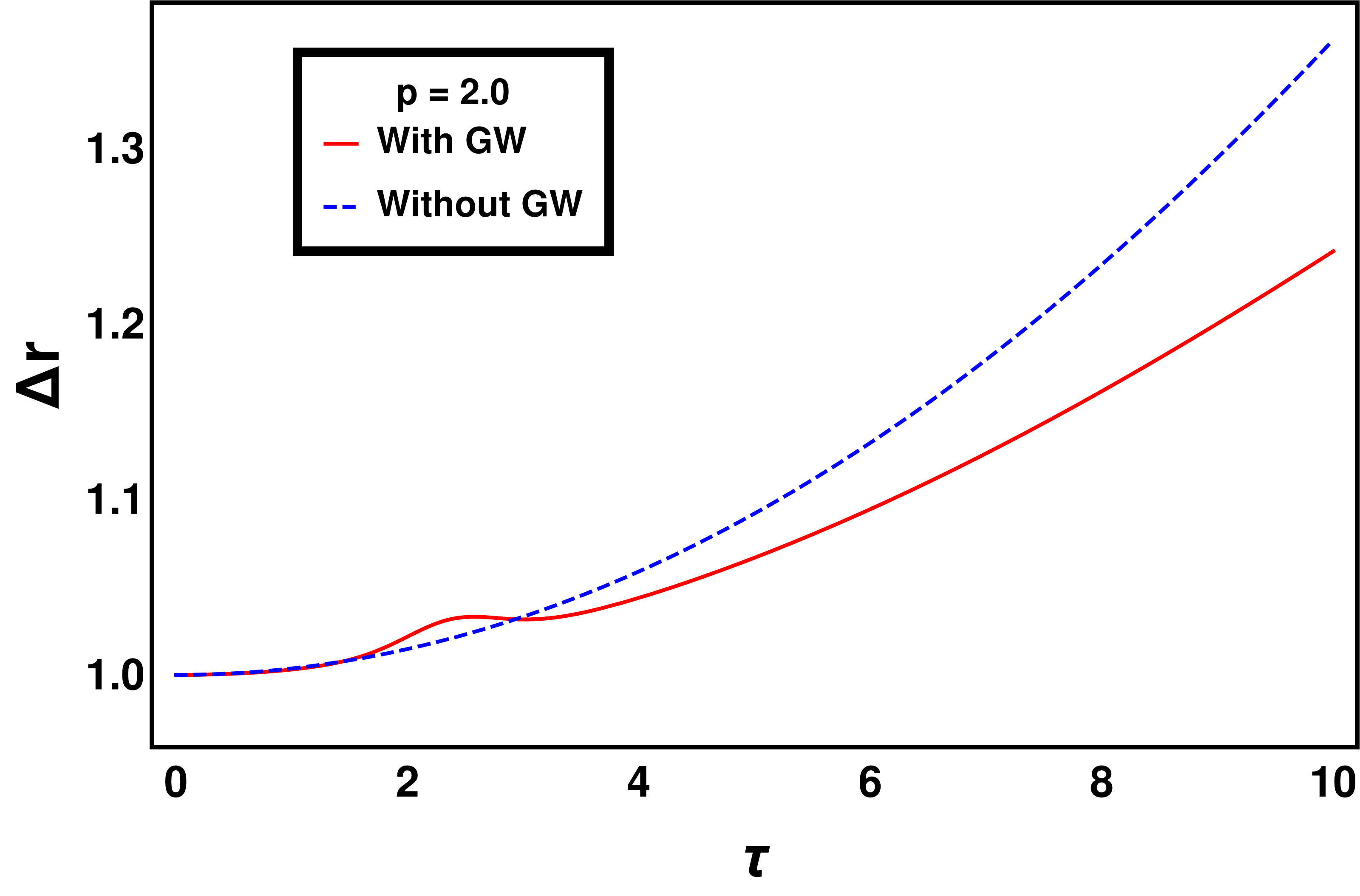}
		 \caption{\centering{\small Variation of the difference between the radial coordinates of the timelike geodesics, namely $\Delta r$, has been depicted with the proper time $\tau$ in the presence as well as in the absence of the GW pulse.}}
		\label{fig:rwgw}
	\end{subfigure}
	\caption{\centering{{\small The variation of $\Delta u$ and $\Delta r$ with the proper time of the geodesics have been presented in the presence and in the absence of GW pulse. The plots explicitly demonstrates that both $\Delta u$ and $\Delta r$ encodes information about the passage of the GW pulse in the past, thus depicts the displacement memory effect.}}}
	\label{fig:Variation_gw}
\end{figure*}
\begin{figure*}[t]
	\centering
	\begin{subfigure}[t]{0.45\textwidth}
		\centering
		\includegraphics[width=\textwidth]{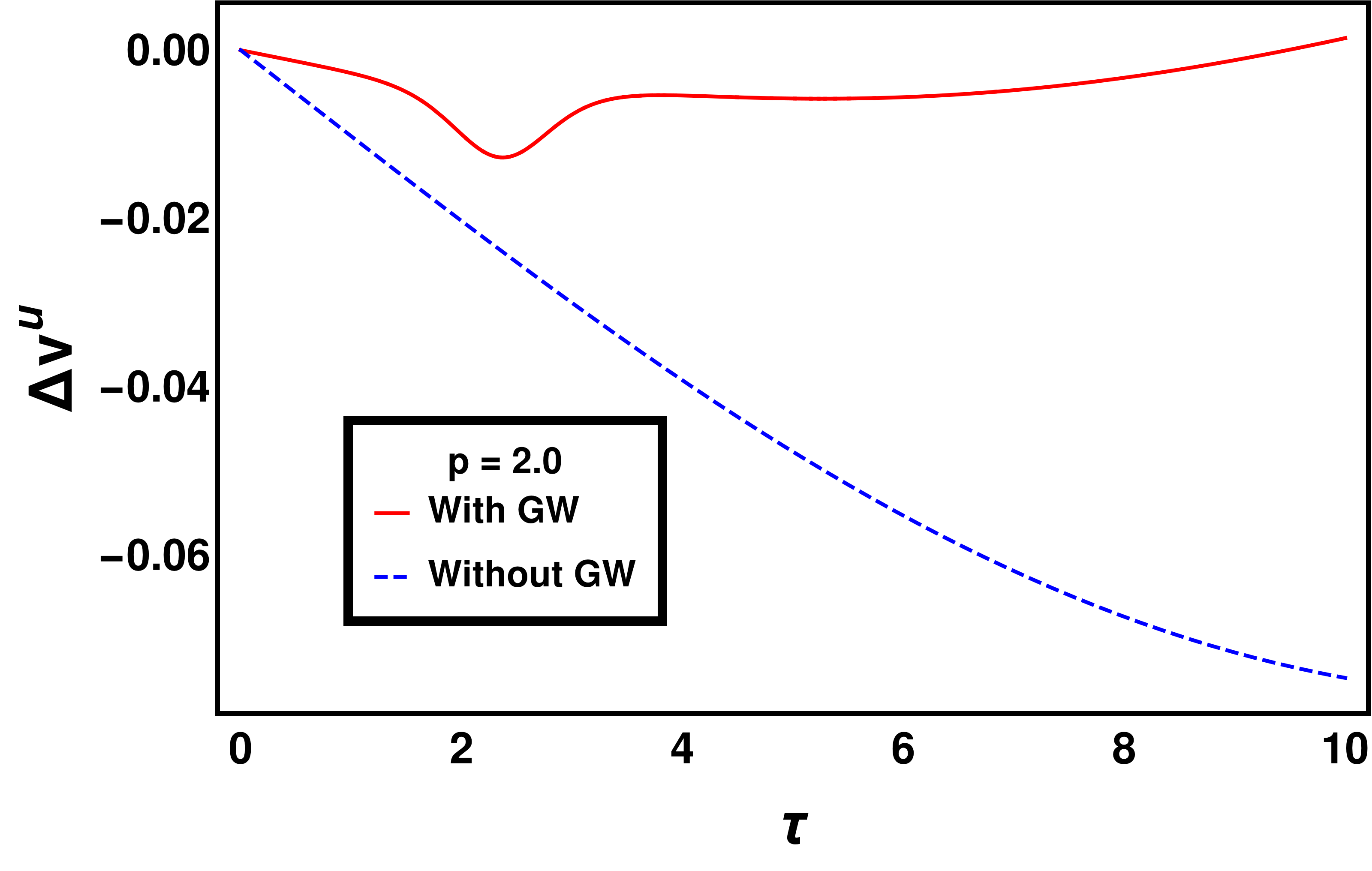}
		 \caption{\centering{\small Variation of the difference between the velocity along null direction $u$ of the two timelike geodesics, have been presented against the proper time $\tau$, in the presence as well as in the absence of the GW pulse.}}
		\label{fig:vugw}
\end{subfigure} \hspace{1cm}
	\begin{subfigure}[t]{0.45\textwidth}
		\centering
		\includegraphics[width=\textwidth]{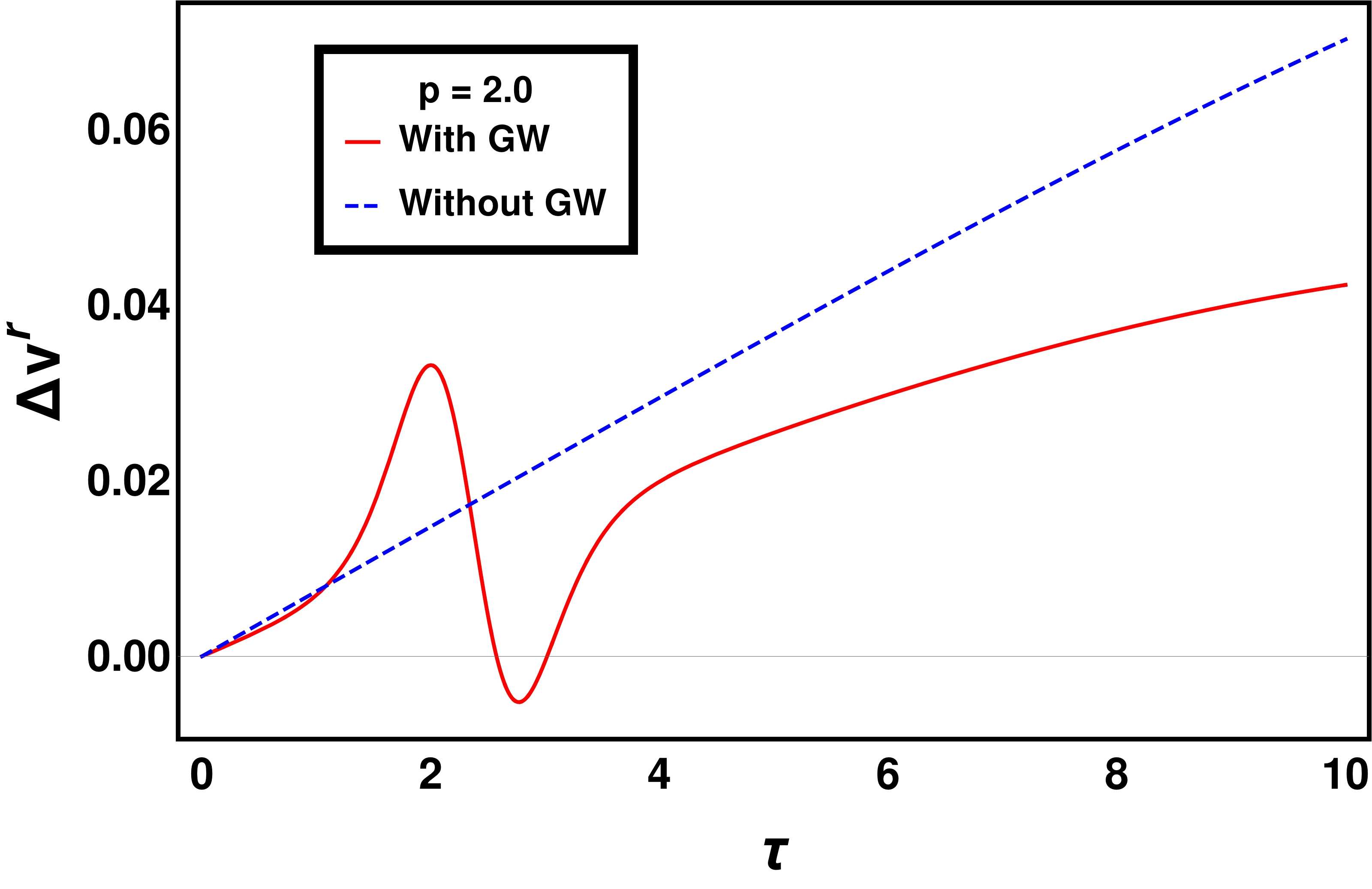}
		 \caption{\centering{\small Variation of the difference between the velocity along the radial direction $r$ of the timelike geodesics has been presented with the proper time $\tau$ with and without the GW pulse.}}
		\label{fig:vrgw}
	\end{subfigure}
	\caption{\centering{{\small We have explicitly depicted that the expressions for $\Delta \dot{u}$ and $\Delta \dot{r}$ deviate significantly in the presence of a GW pulse. This immediately suggests the existence of a velocity memory effect.}}}
	\label{fig:Variation_gw_vel}
\end{figure*}

We have solved the three geodesic equations, presented in \ref{udot} - \ref{fdot} numerically in the symbolic manipulation software {\em Mathematica} and have analysed the solutions extensively. In particular, we have started by considering two neighbouring geodesics in the wormhole background and then have studied the evolution of their coordinate separation in terms of the proper time $\tau$, which is also an affine parameter, see \ref{fig:Variation_du_dr}. Further, we have depicted in \ref{fig:Variation_gw}, how the evolution of these coordinate separations have been affected because of the presence of the GW pulse and also by the presence of extra dimension through non-zero values of $p$. In particular, as \ref{fig:Variation_gw} demonstrates, even after the GW pulse has passed, there is a residual effect which manifests as the GW memory, more specifically, as the displacement memory effect. It turns out that these geodesics also inhibit velocity memory effects, which can be seen from \ref{fig:Variation_gw_vel}.

Let us discuss the above scenario in detail and consider the various boundary conditions imposed on them. First of all, in this work we have considered two neighbouring timelike geodesics, which in the background wormhole geometry would satisfy the following condition on the equatorial plane,
\begin{equation}\label{timelike}
-f(r) \dot u^2 -2 g(r) \dot u \dot r + r^2 \dot \phi^2 - r H(u) \dot \phi^2 = -1~.
\end{equation}
We choose the initial conditions as follows: both the geodesics are chosen to have an initial separation in the radial coordinate $r$, as well as in the null coordinate $u$, however they both start at the same azimuthal coordinate $\phi$. These geodesics have fixed initial values of $\dot{r}$ and $\dot{u}$, while the value of $\dot{\phi}$ will depend on the background geometry through \ref{timelike}. Then we define the following quantities $\Delta u$ and $\Delta r$, as follows,
\begin{eqnarray*}
\Delta u = u({\rm Geodesic~II}) - u(\rm Geodesic~I)~,
\\
\Delta r = r({\rm Geodesic~II}) - r(\rm Geodesic~I)~,
\end{eqnarray*}
where, $u(\rm Geodesic~I)$ and $r(\rm Geodesic ~I)$ corresponds to the coordinates associated with the geodesic I and $u(\rm Geodesic ~II)$ and $r(\rm Geodesic ~II)$ are the coordinates associated with the geodesic II. All of these are obtained by solving the geodesic equations, using appropriate initial conditions, as discussed before. We choose our pulse profile such as $A = 1$ and $u_0 = 5$. The results arising out of the time evolution of $\Delta u$ and $\Delta r$ for different values of wormhole parameter $p$ have been depicted in \ref{fig:Variation_du_dr}.
We consider three cases corresponding to the following values of $p$ --- i) $p = 2.0$, ii) $p = 1.5$ and iii) $p = 1.05$. { The value of the mass is set to $M=3$ for obtaining the plots.} It is clearly seen from these plots that $\Delta u$ is increasing with the decreasing values of $p$ (\ref{fig:kappau}), however, $\Delta r$ is decreasing with the decreasing values of $p$ (\ref{fig:kappar}), i.e., $\Delta u$ and $\Delta r$ have opposite behaviours with change of $p$.  As evident from these figures, the change $\Delta u$ is more pronounced as compared to $\Delta r$. For clarity, in \ref{fig:Variation_du_dr}, the GW pulse is represented by the filled region. The pulse we have shown in the plot is not the exact profile of the GW, rather we have used a scaled-up version of the original profile. The affine parameter interval for which the GW pulse remained significant remains unaltered. We will now discuss the emergence of displacement and velocity memory effect in the wormhole geometry. 

Existence of displacement memory effect is evident from \ref{fig:Variation_gw}, where we have shown that the differences $\Delta u$ and $\Delta r$ between two neighbouring timelike geodesics depends on whether a GW pulse has passed through them or, not. In other words, the values of $\Delta u$ and $\Delta r$ after the GW has passed through does not return to that of the wormhole background and hence provides the displacement memory effect. This memory effect not only depends on the strength of the GW pulse, but more so on the background spacetime geometry. Different choices for the parameter $p$ will lead to different deviations and hence to different memories. In particular, the memory effect can be a potential candidate to test the existence of additional hairs in the background spacetime outside a compact object. In the present scenario, it corresponds to the fact that non-zero $p$ does affect the displacement memory in neighbouring geodesics. 

Finally, in addition to the displacement memory effect, the GW pulse in the wormhole geometry also describes a velocity memory effect, as clear from \ref{fig:Variation_gw_vel}. Both $\Delta \dot{u}$ and $\Delta \dot{r}$ are non-zero and differ from their background values in the presence of the GW pulse. This memory effect also depends on the choice of the wormhole hair $p$ and possibly a combined study of displacement and velocity memory effect will lead to an existential proof of non-zero values of $p$. Therefore, we can conclude that both displacement and velocity memory effects exist in the case of braneworld wormhole and depends crucially on the choice of $p$. This provides another avenue to search for these non-black hole compact objects, which can also hint at the existence of an extra spatial dimension. 


\section{Bondi-Sachs formalism and memory effect from null infinity}\label{null_memory}

Having discussed the displacement and the velocity memory effects from the geodesic analysis in the previous section, we now focus our attention in investigating the memory effect at null infinity \cite{Christodoulou:1991}. Since the effective gravitational field equations on the brane, as presented in \ref{eq:field_equation}, is equivalent to the generalized Brans-Dicke theory, with the Brans-Dicke parameter $\omega$ being also a function of the radion field, we try to reformulate the Bondi-Sachs analysis of the memory effect as prescribed in \cite{Hou:2020JHEP,Hou:2020} appropriately. As in the previous section, here also we assume that the spacetime of interest corresponds to a background wormhole spacetime with a GW pulse passing through it, ultimately reaching the future null infinity and as a consequence modifying the Bondi mass aspect. In particular, we will derive the functional form of the Bondi mass aspect in terms of the wormhole parameters, which in turn are related to the physics of the higher spacetime dimensions through the background wormhole solution. Note that without the GW pulse, there shall be no dynamics associated with the bondi mass of the background wormhole metric, since it depicts a static background.

For this purpose, we have to express the background wormhole geometry in the null coordinates, as presented in \ref{wormhole_line_element}, in the Bondi-Sachs form by an expansion of the same in the inverse powers of the radial coordinate $r$. Such an expansion yields,
\begin{equation}
\begin{split}
ds^2=&-du^2 -2 du dr +r^2 \gamma_{AB} dx^A dx^B 
\\ 
&+\frac{2M}{r(p+1)} du^2- \frac{2Mp}{r(p+1)} du dr + \mathcal{O}(r^{-2})~,
\label{eq:background_whm}
\end{split}
\end{equation}
where, $\gamma_{AB}$ is the metric on the unit-two sphere. The terms in the first line of \ref{eq:background_whm} is the flat metric at future null infinity and the terms in the second line are the desired $(1/r)$ corrections. As evident, the Bondi mass aspect is simply given by $\{M/(p+1)\}$ and is a constant for the background spacetime. This provides the behaviour of the background static wormhole geometry at future null infinity, while the similar result for matter fields will now follow. 

The matter fields consist of two parts, the matter on the four dimensional brane and the radion field. The dependence of the on-brane matter field on the radial coordinate $r$ can be found in \cite{Kar:2015} and it follows that at future null infinity $\mathscr{I}^+$, the matter components fall off faster than $1/r^2$ and hence need not be considered for the calculation. On the other hand, the fall off behaviour of the radion field\footnote{In \ref{eq:scalar_field_isotropic} we had written the scalar field in terms of an isotropic coordinate $\tilde{r}$. As in the asymptotic limit the isotropic coordinate $\tilde{r}$ is equal $r$, one can safely expand the scalar field in terms of $\mathcal{O}(1/r)$.} at future null infinity can be parametrized as: $\Phi_{\rm b}=\Phi_{0\textrm{(b)}}+\Phi_{1\textrm{(b)}}/r+\Phi_{2\textrm{(b)}}/r^2+\cdots$, where `b' denotes that these are contribution from the background wormhole geometry. The term $\Phi_{0\textrm{(b)}}$, which is the leading order term in the asymptotic expansion of the radial field, reads,
\begin{align}
\Phi_{0\textrm{(b)}}&=\bigg(\frac{C_1}{M}\log \frac{p+1}{p-1}+C_4\bigg)^2-1~. \label{eq:Phi_0_radion}
\end{align}
In an identical manner, we obtain the coefficients of $(1/r)$ and $(1/r^{2})$ terms, in the asymptotic expansion of the radion field $\Phi$ as, 
\begin{align}
\Phi_{1\textrm{(b)}}&=-\frac{2C_1}{p+1}\bigg(\frac{C_1}{M}\log \frac{p+1}{p-1}+C_4\bigg) ~,
\label{eq:Phi_1_radion} 
\\
\Phi_{2\textrm{(b)}}&=\frac{C_1^2} {(p+1)^2}+\frac{2Mp C_1}{(p+1)^2}\bigg(\frac{C_1}{M}\log \frac{p+1}{p-1} +C_4\bigg) ~,
\label{eq:Phi_2_radion}
\end{align}
where, $C_1$ and $C_4$ are nonzero constants used to quantify the radion field \citep{Kar:2015}. Note that, these expansion coefficients can be expressed in terms of the parameters of the wormhole spacetime and in particular it depends on $p$. This fact will be used in the later parts of this work.

The above analysis is about the background spacetime, which is definitely non-radiative.  This is because the metric given in \ref{eq:background_whm} has a constant and non-dynamical Bondi mass. This is because there is no loss of {\em news} in absence of any dynamics, as fit for a non-radiative geometry. Thus, the memory effect requires a propagating GW pulse on top of this background geometry, leading to a finite radiative term. Hence, we introduce a GW component to the above wormhole metric and the final result is the following axi-symmetric line element (see \cite{Bondi:1962}), 
\begin{align}\label{eq:final_metric}
ds^2= & -du^2-2dudr  +\bigg(\frac{2M}{r(p+1)}  
+ \frac{2M_{\rm B}(u,\theta)}{r} \bigg)du^2
\nonumber
\\
& -\bigg(\frac{2Mp}{r(p+1)} +\frac{b(u,\theta)}{r}\bigg) du\, dr 
 -2r^2 U(u,r,\theta) du\, d\theta 
\nonumber
\\ 
& + r^2 h_{AB} dx^A dx^B+ \cdots~.  
\end{align}
Note that in the line element presented above, there is a dynamical Bondi mass term $M_{\rm B}(u,\theta)$. This is due to the presence of the gravitational radiation in the background of the wormhole spacetime.

Here, $h_{AB}=\gamma_{AB}+c_{AB}r^{-1}+\mathcal{O}(r^{-2})$ and the terms $b(u,\theta), M_{\rm B}(u,\theta), U(u,r,\theta)$ and $c_{AB}$ arise due to the presence of the GW pulse and can be considered as perturbations over and above the braneworld wormhole metric. Moreover, for the above perturbed metric to be consistent with the gravitational field equations, the radion field should be perturbed as well and the resultant field becomes $\Phi=\Phi_{\rm b}+\Phi_{\rm p}$, where $\Phi_{\rm b}$ denotes the background radion scalar field and $\Phi_{\rm p}$ denotes the perturbed scalar field due to the scalar part of the GW pulse. We assume that the leading order term in $\Phi_{\rm p}$ is $\mathcal{O}(1/r)$, such that the Bondi determinant condition can be expressed as,
\begin{equation}
\det(h_{AB})= (\Phi_{0\textrm{(b)}}/\Phi)^2 \sin^2\theta ~.
\label{eq:det}
\end{equation}
which yields , 
\begin{equation}
c_{AB}= \hat{c}_{AB}-\gamma_{AB}(\Phi_1/\Phi_{0(\textrm{b})})~.
\end{equation}
Here, $\hat{c}_{AB}$ is the pure gravitational part and corresponds to the transverse and traceless degrees of freedom. Also, in the above expression, $\Phi_{1}=\Phi_{1(\textrm{b})}+\Phi_{1(\textrm{p})}$, is the coefficient of the $\mathcal{O}(1/r)$ term in the asymptotic expansion of the radion field $\Phi$. Since there is an additional GW pulse being considered here, it will be described by a tensorial News $N_{AB}$ as well as a scalar News $N$, which are given by,
\begin{align}
-\partial_u \hat{c}_{AB}&=N_{AB}
\nonumber
\\
&=\mathcal{N}_{1} \sech^2 u \, (_{-2}Y^{20})\, \, 
\begin{pmatrix}1 & 0\\ 0 & -\sin^2\theta
\end{pmatrix}~,
\\
N&=\mathcal{N}_{2} \sech^2 u\, (_{-2}Y^{20}) 
\equiv \partial_u \Phi_{1(\textrm{p})}~.
\end{align}
Note that the $\sech^2 u$ behaviour is assumed in order to be consistent with the discussion in the previous section and $N_{AB}$ embodies the gravitational degrees of freedom, while $N$ encodes the scalar degree of freedom. The amplitudes $\mathcal{N}_{1}$ and $\mathcal{N}_{2}$ are such that the GW pulse can be considered as perturbations over the wormhole background. Now, the Bondi determinant condition, along with integration of the above relations over the null coordinate $u$, yields the following change in the Bondi shear $c_{AB}$ as,

\begin{widetext}
\begin{equation}
    \begin{split}
     \triangle c_{AB }&= \triangle \hat{c}_{AB}-\gamma_{AB} \frac{\triangle \Phi_{1(\textrm{p})}}{\Phi_{0(\textrm{b})}}
=-\mathcal{N}_{1} \, (_{-2}Y^{20})\, \, 
\begin{pmatrix}1 & 0\\ 0 & -\sin^2\theta
\end{pmatrix}  \int_{-\infty}^{\infty} \sech^2 u\, du \, \nonumber
\\ 
& -\frac{\mathcal{N}_{2}}{\Phi_{0(\textrm{b})}}(_{-2}Y^{20})\begin{pmatrix}1 & 0\\ 0 & \sin^2\theta
\end{pmatrix}
\int_{-\infty}^{\infty} \sech^2 u\, du 
=-2 \,(_{-2}Y^{20}) \,\begin{bmatrix} \mathcal{N}_{1}+\dfrac{\mathcal{N}_{2}}{\Phi_{0(\textrm{b})}} & 0 \\ 0 & \sin^2\theta \bigg(-\mathcal{N}_{1}+\dfrac{\mathcal{N}_{2}}{\Phi_{0(\textrm{b})}}\bigg) 
    \end{bmatrix} \label{eq:bondi_shear}
    \end{split}
\end{equation}
\end{widetext}


Here, the term $_{-2}Y^{20}$ is the spin-weighted \footnote{Since the system is axisymmetric, the spherical harmonic is chosen in a way such that there is no dependence on $\phi$.} harmonic, having the expression: $_{-2}Y^{20}=(3/4)\sqrt{5/6\pi}~\sin^2\theta$. The above equation, namely \ref{eq:bondi_shear}, shows that the change in the Bondi shear, represented by $\triangle c_{AB }$ is not traceless, with the trace being dependent on the presence of the scalar memory via $\Phi_{0(\textrm{b})}$. Therefore, the existence of a non-zero trace for the Bondi shear will signal the presence of an additional scalar degree of freedom in the system, possibly arising from extra dimensions. Thus, the total gravitational memory in the spacetime has both a tensorial part and a scalar part, the later arising from the radion scalar field of the underlying theory. 
\begin{figure*}
	\centering
	\begin{subfigure}[t]{0.45\textwidth}
		\centering
		\includegraphics[width=\textwidth]{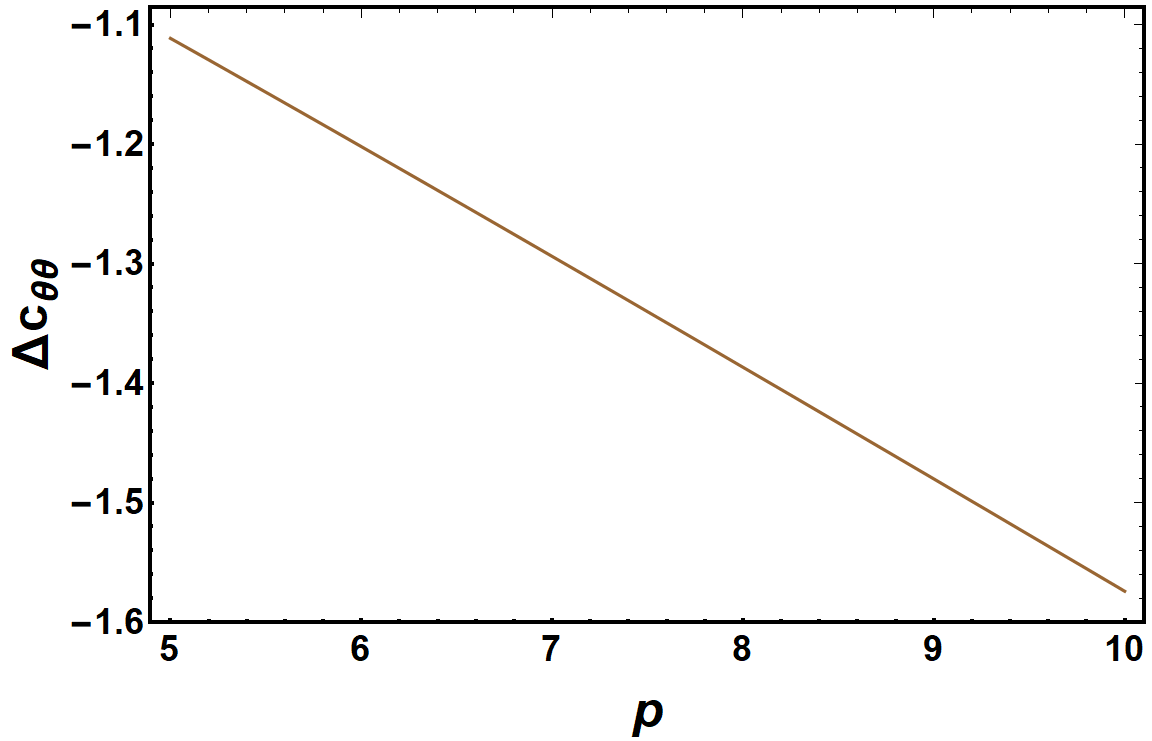}
		 \caption{\centering{\small We have depicted the variation of the Bondi shear with the wormhole hair $p$ on the $\theta=(\pi/2)$ plane.}}
		\label{fig:Variation_kappa}
\end{subfigure} \hspace{1cm}
	\begin{subfigure}[t]{0.45\textwidth}
		\centering
		\includegraphics[width=\textwidth]{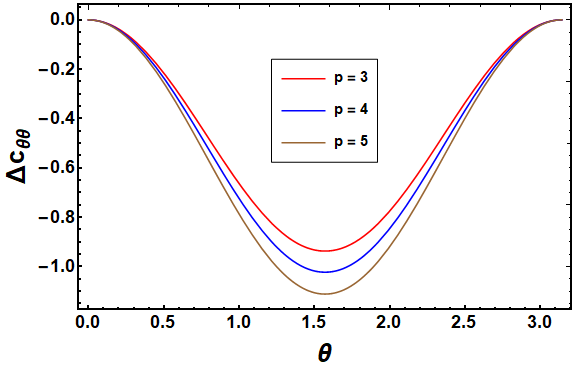}
		 \caption{\centering{\small Response of the change in Bondi shear due to the GW pulse has been presented against $\theta$ for different values of $p$.}}
		\label{fig:Variation_theta_kappa}
	\end{subfigure}
	\caption{\centering{{\small Variation of Bondi shear due to the passage of a GW pulse, also known as the memory tensor has been presented for various choices of the wormhole parameters.}}}
	\label{fig:Variation_kappa_lambda}
\end{figure*}
The effects described above, are clearly depicted in \ref{fig:Variation_kappa_lambda}, where the behaviour of the Bondi shear has been presented with variation in the wormhole parameter $p$. 

Finally, we solve the gravitational field equations order by order, to obtain the change in the Bondi mass aspect of the system, namely $\Delta M_{\rm B}$. For this purpose, we re-express \ref{eq:final_metric} in a convenient form, fit for an axisymmetric system \citep{Bondi:1962}, such that,
\begin{widetext}
\begin{equation}
\begin{split}
ds^2=&-\exp\left[\sigma(u,r,\theta)\right]du^2-2\exp\left[2\beta(u,r,\theta)\right]dudr  -2r^2\exp\left[2\left\{\gamma(u,r,\theta)-\delta(u,r,\theta)\right\}\right]U_1(u,r,\theta)du \, d\theta
\\
&\hskip 2 cm +r^2\exp\left[2\left\{\gamma(u,r,\theta)-\delta(u,r,\theta)\right\}\right]d\theta^2 
+r^2\exp\left[2\left\{-\gamma(u,r,\theta)-\delta(u,r,\theta)\right\}\right] d\phi^2~. \label{eq:bondi_sachs_formalism}
\end{split}
\end{equation}
\end{widetext}
The metric functions $\sigma$, $\beta$, $U_1$, $\gamma$ and $\delta$, appearing in the above expression, are all expanded in the inverse powers of the radial coordinate $r$, yielding, 
\begin{gather}
\sigma(u,r,\theta)=\frac{\sigma_1(u,\theta)}{r}+\frac{\sigma_2(u,\theta)}{r^2}+\mathcal{O}(r^{-3})
\\
\beta(u,r,\theta)=\frac{\beta_1(u,\theta)}{r}+\frac{\beta_2(u,\theta)}{r^2}+\mathcal{O}(r^{-3})
\\
U_1(u,r,\theta)=\frac{U_{11}(u,\theta)}{r}+\frac{U_{12}(u,\theta)}{r^2}+\mathcal{O}(r^{-3})
\\
\gamma(u,r,\theta)=\frac{\gamma_1(u,\theta)}{r}+\frac{\gamma_2(u,\theta)}{r^2}+\mathcal{O}(r^{-3})
\\
\delta(u,r,\theta)=\frac{\delta_1(u,\theta)}{r}+\frac{\delta_2(u,\theta)}{r^2}+\mathcal{O}(r^{-3})
\end{gather}
These expansions must be compared with the Bondi-Sachs form of the wormhole metric with a GW pulse, as presented in \ref{eq:final_metric}, from which we find the following correspondence,
\begin{eqnarray}
\sigma_1(u,\theta)&=&-\frac{2M}{p+1}-2M_{\rm B}(u,\theta)~,
\nonumber 
\\
\beta_1(u,\theta)&=&\frac{M\,p}{p+1}+\frac{b(u,\theta)}{2}~, 
\nonumber 
\\
\delta_1(u,\theta)&=&\frac{\Phi_1}{2\Phi_{0(\textrm{b})}}~,
\nonumber 
\\
\gamma_1(u,\theta)&=&\frac{\hat{c}_{\theta\theta}}{2}\equiv \frac{\hat{c}(u,\theta)}{2}~,
\nonumber  
\\
\delta_2(u,\theta)&=&U(u,r,\theta)
\nonumber
\\
&=&\dfrac{\Phi_2 U_1(u,r,\theta)}{2\Phi_{0(\textrm{b})}} \exp\left[2\left\{\gamma(u,r,\theta)-\delta(u,r,\theta)\right\}\right]~.
\nonumber
\\
\end{eqnarray}
\begin{figure*}
	\centering
		\includegraphics[scale=0.3]{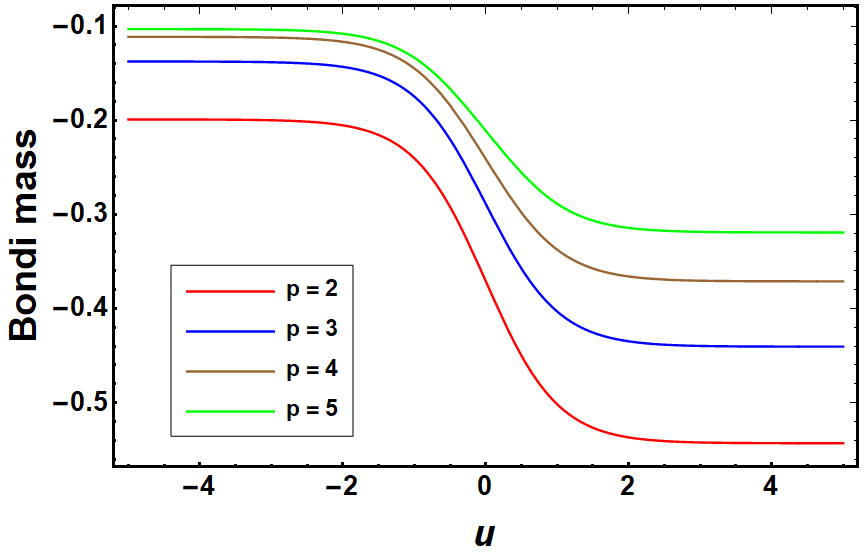}
		 \caption{\centering{Variation of the change in the Bondi mass due to GW pulse has been presented with the null coordinate $u$, for different choices of $p$.}}
	\label{fig:bondi_mass}
	\end{figure*}
To proceed further, we use the fact that the field equations in Bondi-Sachs formalism follows a nested pattern. One starts with an initial data, prescribed in terms of the functions $\hat{c}_{\theta \theta}(u,\theta)$ and $\Phi_{1(\textrm{p})}(u,\theta)$ at some value of $u$. Subsequently, the hypersurface equations $G^u\,_r$ and $G^u\,_\theta$, respectively yields, \footnote{The following field equations are solved using the RGTC package in the symbolic manipulation software {\em Mathematica}.}
\begin{align}
\mathcal{O}(r^{-1})&:\hspace{0.2cm} U_{11}=0~,  
\\
\mathcal{O}(r^{-2})&:\hspace{0.2cm} U_{12}(u,\theta)=-\frac{\partial_{\theta}\hat{c}(u,\theta)}{2}-\hat{c}(u,\theta) \cot\theta~,
\end{align}
and finally, at order $\mathcal{O}(r^{-3})$, we obtain,
\begin{equation}
\beta_1=-\frac{\Phi_1(u,\theta)}{2\Phi_{0(\textrm{b})}}~;\qquad
b(u,\theta)=-\frac{\Phi_1(u,\theta)}{\Phi_{0(\textrm{b})}}-\frac{2M\,p}{p+1}~.
\end{equation}
Similarly, the hypersurface equation in the $G^u\,_u$ component of Einstein's equations yields an identity at the lowest order of $\mathcal{O}(r^{-2})$, while the supplementary equation for the component $G^r\,_u$ at $\mathcal{O}(r^{-2})$ yields the change in the Bondi mass aspect of the system,
\begin{equation}
\begin{split}
\sigma_1,_u= & \frac{\left(\partial_{u}\Phi_1\right)^2}{2\Phi_0^2}-\frac{3\left(\partial_{u}\Phi_1\right)^2}{2\Phi_0(1+\Phi_0)}-\frac{\Phi_1\partial_{u}^{2}\Phi_1}{\Phi_0^2}
\\
&-\frac{\partial_{u}\Phi_1,}{\Phi_0}+\frac{\left(\partial_{u}\hat{c}\right)^2}{2} +\partial_{u}\hat{c}-\frac{3}{2}\partial_{u}\partial_{\theta}\hat{c}\cot\theta-\frac{1}{2}\partial_{u}\partial_{\theta}^{2}\hat{c}~.
\label{eq:bondi_mass_eqn}
\end{split}
\end{equation}
Integrating the above expression, we obtain the effective Bondi mass of the system to yield,
\begin{widetext}
\begin{equation}
\begin{split}
M_{\rm B}= & -\frac{m}{p+1}-\frac{1}{2}\bigg[\frac{a^2Y^2}{2}+b^2Y^2\bigg\{\frac{1}{2\Phi_0^2}-\frac{3}{2\Phi_0(1+\Phi_0)}\bigg\}\bigg]
\times\bigg(\frac{2}{3}\tanh u +\frac{1}{3}\sech^2u\tanh u\bigg)
\\
& -\frac{1}{3\Phi_0^2}b^2Y^2\tanh^3u  +\frac{1}{2}\tanh u\bigg(\frac{bY}{\Phi_0}+aY+\frac{3}{2}aY,_\theta \cot\theta+aY,_{\theta\theta}\bigg)~. \label{eq:bondi_mass}
\end{split}
\end{equation}
\end{widetext}
The evolution of the Bondi mass with variation in $p$ has been depicted in \ref{fig:bondi_mass}. In the plot we find that the drop in the Bondi mass is higher as the value of $p$ decreases. 

\section{Conclusions}

In this article, we have explored certain aspects of the GW memory in a wormhole background on the brane. The reason for considering the presence of extra spacetime dimension is two-fold --- i) The on-brane gravity theory is a quasi scalar-tensor theory, with the scalar field capturing the imprints of the spatial extra dimension and hence any information about the scalar hair will translate into possible inputs for the extra dimensions.  ii) In this class the wormhole geometry is traversable and can be sustained without invoking any exotic matter fields. In this manner we arrive at a possible non-black hole compact object without using exotic matter field and hence it provides interesting viable alternative to the standard black hole paradigm in GR. We would like to mention that, besides the wormhole solution considered here, there are other wormhole solutions, e.g., in the context of Scalar-coupled Einstein-Gauss-Bonnet Gravity \cite{Antoniou:2019awm}, where also the matter field is not exotic. It will be interesting to study the stability and hence the memory effect in such wormhole backgrounds as well.

We, at first, have briefly reviewed the geometry of the wormhole spacetime and how the presence of extra dimension helps in constructing a traversable wormhole without any exotic matter. Then we have explored the displacement and velocity memory effects by analysing neighbouring geodesics in the wormhole background in the presence of a localised GW pulse. We have shown explicitly, how the geodesic separation evolves before and after the passage of the pulse. This explicitly establishes the existence of both displacement and velocity memory effect. In addition, these memory effects depend crucially on the hairs of this wormhole solution and hence differs from the corresponding memory effect in Schwarzschild spacetime. Therefore, memory effects are indeed future pointers towards exploring existence of non-black hole compact objects in our universe, which in this context can also be related to the existence of extra spatial dimensions.  

Having observed that indeed memory effect exists and depends on the details of the wormhole geometry, we subsequently proceeded to study the memory effect using symmetries at null infinity using the Bondi-Sachs formalism. For this purpose, we have expressed the spacetime metric of the wormhole geometry using the Bondi coordinate system and have expanded the radion field in inverse powers of the radial coordinate $r$. Considering a GW perturbation in the system that satisfies the Bondi gauge conditions, we have computed the Bondi shear as well as the Bondi mass aspect and hence observed that these also give rise to the memory effect. Again, the memory depends on the wormhole parameter ($p$) through the leading order contribution from the radial field. 

Since the braneworld scenario, considered in the present context, very much resembles a scalar-tensor theory of gravity, we use the formalism given in \cite{Hou:2020JHEP}, and show how the variation in the Bondi mass aspect, related to the memory effect, depend explicitly on the wormhole parameters. The same conclusion holds true for geodesic memory effects as well. This variation of the Bondi mass aspect is different from the black hole scenario and can possible be probed using the future GW detectors.  Moreover, generalization of the present result for astrophysically relevant cases of rotating wormholes will be of significant interest. Besides, this work can also be used as a prospect to investigate supertranslations and soft hair implants on the throat of a wormhole geometry (analogous study for black hole horizons can be found in \cite{Hawking:2016} and Rindler horizons in \cite{Kolekar:2017}). These issues we wish to study in the future. 

The memory effect encoded in the Bondi mass aspect is computed by solving the gravitational field equations of the theory at orders of $\mathcal{O}(1/r)$. Decomposing the field equations into Hypersurface and Supplementary equations, we have determined the change in the Bondi mass of the system analytically.  We find that the variation in the value of $p$ produces a change in the evolution of the Bondi mass.  This shows that it will be easier to decipher whether the central compact object has a non-zero value of $p$, since presence of a non-zero $p$ will modify the memory significantly. Therefore, as and when the memory effect can be detected in the future GW detectors, it will possibly tell us about the existence of non-black hole compact objects and one can independently verify if the braneworld wormhole is a viable scenario.   

There are several possible future extensions of the present work, first of all a complete analysis of the Bondi-Sachs formalism without assuming axisymmetry will be an interesting and important extension. Moreover, studying memory effect for rotating wormholes and other rotating compact objects, which are more relevant astrophysically, will be another interesting avenue to explore. Finally, studying GW memory effects for other black hole mimicker spacetimes, e.g., fuzzballs will be of significant interest.  


\section*{Acknowledgements}

I.C. acknowledges the University Grants Commission (UGC), Government of India, for providing financial assistance through a senior research
fellowship (reference ID: 523711). Research of S.C. is funded by the INSPIRE Faculty fellowship from DST, Government of India (Reg. No. DST/INSPIRE/04/2018/000893) and by the Start- Up Research Grant from SERB, DST, Government of India (Reg. No. SRG/2020/000409). S.C. further thanks the Albert-Einstein Institute, where a part of this work was carried out and the Max-Planck Society for providing the Max-Planck-India Mobility Grant.  

\bibliographystyle{apsrev4-2}
\bibliography{mybibliography_whm}

\begin{thebibliography}{88}%
\makeatletter
\providecommand \@ifxundefined [1]{%
 \@ifx{#1\undefined}
}%
\providecommand \@ifnum [1]{%
 \ifnum #1\expandafter \@firstoftwo
 \else \expandafter \@secondoftwo
 \fi
}%
\providecommand \@ifx [1]{%
 \ifx #1\expandafter \@firstoftwo
 \else \expandafter \@secondoftwo
 \fi
}%
\providecommand \natexlab [1]{#1}%
\providecommand \enquote  [1]{``#1''}%
\providecommand \bibnamefont  [1]{#1}%
\providecommand \bibfnamefont [1]{#1}%
\providecommand \citenamefont [1]{#1}%
\providecommand \href@noop [0]{\@secondoftwo}%
\providecommand \href [0]{\begingroup \@sanitize@url \@href}%
\providecommand \@href[1]{\@@startlink{#1}\@@href}%
\providecommand \@@href[1]{\endgroup#1\@@endlink}%
\providecommand \@sanitize@url [0]{\catcode `\\12\catcode `\$12\catcode
  `\&12\catcode `\#12\catcode `\^12\catcode `\_12\catcode `\%12\relax}%
\providecommand \@@startlink[1]{}%
\providecommand \@@endlink[0]{}%
\providecommand \url  [0]{\begingroup\@sanitize@url \@url }%
\providecommand \@url [1]{\endgroup\@href {#1}{\urlprefix }}%
\providecommand \urlprefix  [0]{URL }%
\providecommand \Eprint [0]{\href }%
\providecommand \doibase [0]{https://doi.org/}%
\providecommand \selectlanguage [0]{\@gobble}%
\providecommand \bibinfo  [0]{\@secondoftwo}%
\providecommand \bibfield  [0]{\@secondoftwo}%
\providecommand \translation [1]{[#1]}%
\providecommand \BibitemOpen [0]{}%
\providecommand \bibitemStop [0]{}%
\providecommand \bibitemNoStop [0]{.\EOS\space}%
\providecommand \EOS [0]{\spacefactor3000\relax}%
\providecommand \BibitemShut  [1]{\csname bibitem#1\endcsname}%
\let\auto@bib@innerbib\@empty
\bibitem [{\citenamefont {Abbott}\ \emph {et~al.}(2016)\citenamefont {Abbott}
  \emph {et~al.}}]{Abbott:2016}%
  \BibitemOpen
  \bibfield  {author} {\bibinfo {author} {\bibfnamefont {B.}~\bibnamefont
  {Abbott}} \emph {et~al.} (\bibinfo {collaboration} {LIGO Scientific,
  Virgo}),\ }\href {https://doi.org/10.1103/PhysRevLett.116.061102} {\bibfield
  {journal} {\bibinfo  {journal} {Phys. Rev. Lett.}\ }\textbf {\bibinfo
  {volume} {116}},\ \bibinfo {pages} {061102} (\bibinfo {year}
  {2016})}\BibitemShut {NoStop}%
\bibitem [{\citenamefont {Abbott}\ \emph {et~al.}(2017)\citenamefont {Abbott}
  \emph {et~al.}}]{Abbott:2017}%
  \BibitemOpen
  \bibfield  {author} {\bibinfo {author} {\bibfnamefont {B.~P.}\ \bibnamefont
  {Abbott}} \emph {et~al.} (\bibinfo {collaboration} {LIGO Scientific,
  Virgo}),\ }\href {https://doi.org/10.1103/PhysRevLett.119.161101} {\bibfield
  {journal} {\bibinfo  {journal} {Phys. Rev. Lett.}\ }\textbf {\bibinfo
  {volume} {119}},\ \bibinfo {pages} {161101} (\bibinfo {year} {2017})},\
  \Eprint {https://arxiv.org/abs/1710.05832} {arXiv:1710.05832 [gr-qc]}
  \BibitemShut {NoStop}%
\bibitem [{\citenamefont {Akiyama}\ \emph
  {et~al.}(2019{\natexlab{a}})\citenamefont {Akiyama} \emph {et~al.}}]{EHT:1}%
  \BibitemOpen
  \bibfield  {author} {\bibinfo {author} {\bibfnamefont {K.}~\bibnamefont
  {Akiyama}} \emph {et~al.} (\bibinfo {collaboration} {Event Horizon
  Telescope}),\ }\href {https://doi.org/10.3847/2041-8213/ab0ec7} {\bibfield
  {journal} {\bibinfo  {journal} {Astrophys. J. Lett.}\ }\textbf {\bibinfo
  {volume} {875}},\ \bibinfo {pages} {L1} (\bibinfo {year}
  {2019}{\natexlab{a}})},\ \Eprint {https://arxiv.org/abs/1906.11238}
  {arXiv:1906.11238 [astro-ph.GA]} \BibitemShut {NoStop}%
\bibitem [{\citenamefont {Akiyama}\ \emph
  {et~al.}(2019{\natexlab{b}})\citenamefont {Akiyama} \emph {et~al.}}]{EHT:2}%
  \BibitemOpen
  \bibfield  {author} {\bibinfo {author} {\bibfnamefont {K.}~\bibnamefont
  {Akiyama}} \emph {et~al.} (\bibinfo {collaboration} {Event Horizon
  Telescope}),\ }\href {https://doi.org/10.3847/2041-8213/ab0c96} {\bibfield
  {journal} {\bibinfo  {journal} {Astrophys. J. Lett.}\ }\textbf {\bibinfo
  {volume} {875}},\ \bibinfo {pages} {L2} (\bibinfo {year}
  {2019}{\natexlab{b}})},\ \Eprint {https://arxiv.org/abs/1906.11239}
  {arXiv:1906.11239 [astro-ph.IM]} \BibitemShut {NoStop}%
\bibitem [{\citenamefont {Akiyama}\ \emph
  {et~al.}(2019{\natexlab{c}})\citenamefont {Akiyama} \emph {et~al.}}]{EHT:3}%
  \BibitemOpen
  \bibfield  {author} {\bibinfo {author} {\bibfnamefont {K.}~\bibnamefont
  {Akiyama}} \emph {et~al.} (\bibinfo {collaboration} {Event Horizon
  Telescope}),\ }\href {https://doi.org/10.3847/2041-8213/ab0c57} {\bibfield
  {journal} {\bibinfo  {journal} {Astrophys. J. Lett.}\ }\textbf {\bibinfo
  {volume} {875}},\ \bibinfo {pages} {L3} (\bibinfo {year}
  {2019}{\natexlab{c}})},\ \Eprint {https://arxiv.org/abs/1906.11240}
  {arXiv:1906.11240 [astro-ph.GA]} \BibitemShut {NoStop}%
\bibitem [{\citenamefont {Akiyama}\ \emph
  {et~al.}(2019{\natexlab{d}})\citenamefont {Akiyama} \emph {et~al.}}]{EHT:4}%
  \BibitemOpen
  \bibfield  {author} {\bibinfo {author} {\bibfnamefont {K.}~\bibnamefont
  {Akiyama}} \emph {et~al.} (\bibinfo {collaboration} {Event Horizon
  Telescope}),\ }\href {https://doi.org/10.3847/2041-8213/ab0e85} {\bibfield
  {journal} {\bibinfo  {journal} {Astrophys. J. Lett.}\ }\textbf {\bibinfo
  {volume} {875}},\ \bibinfo {pages} {L4} (\bibinfo {year}
  {2019}{\natexlab{d}})},\ \Eprint {https://arxiv.org/abs/1906.11241}
  {arXiv:1906.11241 [astro-ph.GA]} \BibitemShut {NoStop}%
\bibitem [{\citenamefont {Akiyama}\ \emph
  {et~al.}(2019{\natexlab{e}})\citenamefont {Akiyama} \emph {et~al.}}]{EHT:5}%
  \BibitemOpen
  \bibfield  {author} {\bibinfo {author} {\bibfnamefont {K.}~\bibnamefont
  {Akiyama}} \emph {et~al.} (\bibinfo {collaboration} {Event Horizon
  Telescope}),\ }\href {https://doi.org/10.3847/2041-8213/ab0f43} {\bibfield
  {journal} {\bibinfo  {journal} {Astrophys. J. Lett.}\ }\textbf {\bibinfo
  {volume} {875}},\ \bibinfo {pages} {L5} (\bibinfo {year}
  {2019}{\natexlab{e}})},\ \Eprint {https://arxiv.org/abs/1906.11242}
  {arXiv:1906.11242 [astro-ph.GA]} \BibitemShut {NoStop}%
\bibitem [{\citenamefont {Akiyama}\ \emph
  {et~al.}(2022{\natexlab{a}})\citenamefont {Akiyama} \emph {et~al.}}]{Sgr:1}%
  \BibitemOpen
  \bibfield  {author} {\bibinfo {author} {\bibfnamefont {K.}~\bibnamefont
  {Akiyama}} \emph {et~al.} (\bibinfo {collaboration} {Event Horizon
  Telescope}),\ }\href {https://doi.org/10.3847/2041-8213/ac6674} {\bibfield
  {journal} {\bibinfo  {journal} {Astrophys. J. Lett.}\ }\textbf {\bibinfo
  {volume} {930}},\ \bibinfo {pages} {L12} (\bibinfo {year}
  {2022}{\natexlab{a}})}\BibitemShut {NoStop}%
\bibitem [{\citenamefont {Akiyama}\ \emph
  {et~al.}(2022{\natexlab{b}})\citenamefont {Akiyama} \emph {et~al.}}]{Sgr:6}%
  \BibitemOpen
  \bibfield  {author} {\bibinfo {author} {\bibfnamefont {K.}~\bibnamefont
  {Akiyama}} \emph {et~al.} (\bibinfo {collaboration} {Event Horizon
  Telescope}),\ }\href {https://doi.org/10.3847/2041-8213/ac6756} {\bibfield
  {journal} {\bibinfo  {journal} {Astrophys. J. Lett.}\ }\textbf {\bibinfo
  {volume} {930}},\ \bibinfo {pages} {L17} (\bibinfo {year}
  {2022}{\natexlab{b}})}\BibitemShut {NoStop}%
\bibitem [{\citenamefont {Yunes}\ and\ \citenamefont
  {Siemens}(2013)}]{Yunes:2013}%
  \BibitemOpen
  \bibfield  {author} {\bibinfo {author} {\bibfnamefont {N.}~\bibnamefont
  {Yunes}}\ and\ \bibinfo {author} {\bibfnamefont {X.}~\bibnamefont
  {Siemens}},\ }\href {https://doi.org/10.12942/lrr-2013-9} {\bibfield
  {journal} {\bibinfo  {journal} {Living Rev. Rel.}\ }\textbf {\bibinfo
  {volume} {16}},\ \bibinfo {pages} {9} (\bibinfo {year} {2013})},\ \Eprint
  {https://arxiv.org/abs/1304.3473} {arXiv:1304.3473 [gr-qc]} \BibitemShut
  {NoStop}%
\bibitem [{\citenamefont {{The LIGO Scientific Collaboration}}\ \emph
  {et~al.}(2021)\citenamefont {{The LIGO Scientific Collaboration}},
  \citenamefont {{The Virgo Collaboration}}, \citenamefont {{The KAGRA
  Collaboration}},\ and\ \citenamefont {Abbott}}]{RAbbott_1:2021}%
  \BibitemOpen
  \bibfield  {author} {\bibinfo {author} {\bibnamefont {{The LIGO Scientific
  Collaboration}}}, \bibinfo {author} {\bibnamefont {{The Virgo
  Collaboration}}}, \bibinfo {author} {\bibnamefont {{The KAGRA
  Collaboration}}},\ and\ \bibinfo {author} {\bibfnamefont {R.~e.~a.}\
  \bibnamefont {Abbott}},\ }\href {https://doi.org/10.48550/ARXIV.2112.06861}
  {\bibinfo {title} {arxiv:2112.06861}} (\bibinfo {year} {2021})\BibitemShut
  {NoStop}%
\bibitem [{\citenamefont {Krishnendu}\ and\ \citenamefont
  {Ohme}(2021)}]{Krishnendu:2021}%
  \BibitemOpen
  \bibfield  {author} {\bibinfo {author} {\bibfnamefont {N.~V.}\ \bibnamefont
  {Krishnendu}}\ and\ \bibinfo {author} {\bibfnamefont {F.}~\bibnamefont
  {Ohme}},\ }\href {https://doi.org/10.3390/universe7120497} {\bibfield
  {journal} {\bibinfo  {journal} {Universe}\ }\textbf {\bibinfo {volume} {7}},\
  \bibinfo {pages} {497} (\bibinfo {year} {2021})},\ \Eprint
  {https://arxiv.org/abs/2201.05418} {arXiv:2201.05418 [gr-qc]} \BibitemShut
  {NoStop}%
\bibitem [{\citenamefont {Johannsen}\ \emph {et~al.}(2016)\citenamefont
  {Johannsen}, \citenamefont {Broderick}, \citenamefont {Plewa}, \citenamefont
  {Chatzopoulos}, \citenamefont {Doeleman}, \citenamefont {Eisenhauer},
  \citenamefont {Fish}, \citenamefont {Genzel}, \citenamefont {Gerhard},\ and\
  \citenamefont {Johnson}}]{Johannsen:2015}%
  \BibitemOpen
  \bibfield  {author} {\bibinfo {author} {\bibfnamefont {T.}~\bibnamefont
  {Johannsen}}, \bibinfo {author} {\bibfnamefont {A.~E.}\ \bibnamefont
  {Broderick}}, \bibinfo {author} {\bibfnamefont {P.~M.}\ \bibnamefont
  {Plewa}}, \bibinfo {author} {\bibfnamefont {S.}~\bibnamefont {Chatzopoulos}},
  \bibinfo {author} {\bibfnamefont {S.~S.}\ \bibnamefont {Doeleman}}, \bibinfo
  {author} {\bibfnamefont {F.}~\bibnamefont {Eisenhauer}}, \bibinfo {author}
  {\bibfnamefont {V.~L.}\ \bibnamefont {Fish}}, \bibinfo {author}
  {\bibfnamefont {R.}~\bibnamefont {Genzel}}, \bibinfo {author} {\bibfnamefont
  {O.}~\bibnamefont {Gerhard}},\ and\ \bibinfo {author} {\bibfnamefont {M.~D.}\
  \bibnamefont {Johnson}},\ }\href
  {https://doi.org/10.1103/PhysRevLett.116.031101} {\bibfield  {journal}
  {\bibinfo  {journal} {Phys. Rev. Lett.}\ }\textbf {\bibinfo {volume} {116}},\
  \bibinfo {pages} {031101} (\bibinfo {year} {2016})},\ \Eprint
  {https://arxiv.org/abs/1512.02640} {arXiv:1512.02640 [astro-ph.GA]}
  \BibitemShut {NoStop}%
\bibitem [{\citenamefont {Ayzenberg}\ and\ \citenamefont
  {Yunes}(2018)}]{Ayzenberg:2018}%
  \BibitemOpen
  \bibfield  {author} {\bibinfo {author} {\bibfnamefont {D.}~\bibnamefont
  {Ayzenberg}}\ and\ \bibinfo {author} {\bibfnamefont {N.}~\bibnamefont
  {Yunes}},\ }\href {https://doi.org/10.1088/1361-6382/aae87b} {\bibfield
  {journal} {\bibinfo  {journal} {Class. Quant. Grav.}\ }\textbf {\bibinfo
  {volume} {35}},\ \bibinfo {pages} {235002} (\bibinfo {year} {2018})},\
  \Eprint {https://arxiv.org/abs/1807.08422} {arXiv:1807.08422 [gr-qc]}
  \BibitemShut {NoStop}%
\bibitem [{\citenamefont {Psaltis}(2019)}]{Psaltis:2018}%
  \BibitemOpen
  \bibfield  {author} {\bibinfo {author} {\bibfnamefont {D.}~\bibnamefont
  {Psaltis}},\ }\href {https://doi.org/10.1007/s10714-019-2611-5} {\bibfield
  {journal} {\bibinfo  {journal} {Gen. Rel. Grav.}\ }\textbf {\bibinfo {volume}
  {51}},\ \bibinfo {pages} {137} (\bibinfo {year} {2019})},\ \Eprint
  {https://arxiv.org/abs/1806.09740} {arXiv:1806.09740 [astro-ph.HE]}
  \BibitemShut {NoStop}%
\bibitem [{\citenamefont {Banerjee}\ \emph
  {et~al.}(2020{\natexlab{a}})\citenamefont {Banerjee}, \citenamefont
  {Chakraborty},\ and\ \citenamefont {SenGupta}}]{Banerjee:2019nnj}%
  \BibitemOpen
  \bibfield  {author} {\bibinfo {author} {\bibfnamefont {I.}~\bibnamefont
  {Banerjee}}, \bibinfo {author} {\bibfnamefont {S.}~\bibnamefont
  {Chakraborty}},\ and\ \bibinfo {author} {\bibfnamefont {S.}~\bibnamefont
  {SenGupta}},\ }\href {https://doi.org/10.1103/PhysRevD.101.041301} {\bibfield
   {journal} {\bibinfo  {journal} {Phys. Rev. D}\ }\textbf {\bibinfo {volume}
  {101}},\ \bibinfo {pages} {041301} (\bibinfo {year} {2020}{\natexlab{a}})},\
  \Eprint {https://arxiv.org/abs/1909.09385} {arXiv:1909.09385 [gr-qc]}
  \BibitemShut {NoStop}%
\bibitem [{\citenamefont {Chakraborty}\ \emph {et~al.}(2022)\citenamefont
  {Chakraborty}, \citenamefont {Maggio}, \citenamefont {Mazumdar},\ and\
  \citenamefont {Pani}}]{Chakraborty:2022zlq}%
  \BibitemOpen
  \bibfield  {author} {\bibinfo {author} {\bibfnamefont {S.}~\bibnamefont
  {Chakraborty}}, \bibinfo {author} {\bibfnamefont {E.}~\bibnamefont {Maggio}},
  \bibinfo {author} {\bibfnamefont {A.}~\bibnamefont {Mazumdar}},\ and\
  \bibinfo {author} {\bibfnamefont {P.}~\bibnamefont {Pani}},\ }\href@noop {}
  {\  (\bibinfo {year} {2022})},\ \Eprint {https://arxiv.org/abs/2202.09111}
  {arXiv:2202.09111 [gr-qc]} \BibitemShut {NoStop}%
\bibitem [{\citenamefont {Mishra}\ \emph {et~al.}(2019)\citenamefont {Mishra},
  \citenamefont {Chakraborty},\ and\ \citenamefont {Sarkar}}]{Mishra:2019trb}%
  \BibitemOpen
  \bibfield  {author} {\bibinfo {author} {\bibfnamefont {A.~K.}\ \bibnamefont
  {Mishra}}, \bibinfo {author} {\bibfnamefont {S.}~\bibnamefont
  {Chakraborty}},\ and\ \bibinfo {author} {\bibfnamefont {S.}~\bibnamefont
  {Sarkar}},\ }\href {https://doi.org/10.1103/PhysRevD.99.104080} {\bibfield
  {journal} {\bibinfo  {journal} {Phys. Rev. D}\ }\textbf {\bibinfo {volume}
  {99}},\ \bibinfo {pages} {104080} (\bibinfo {year} {2019})},\ \Eprint
  {https://arxiv.org/abs/1903.06376} {arXiv:1903.06376 [gr-qc]} \BibitemShut
  {NoStop}%
\bibitem [{\citenamefont {Mathur}(2005)}]{Mathur:2005}%
  \BibitemOpen
  \bibfield  {author} {\bibinfo {author} {\bibfnamefont {S.~D.}\ \bibnamefont
  {Mathur}},\ }\href {https://doi.org/10.1002/prop.200410203} {\bibfield
  {journal} {\bibinfo  {journal} {Fortsch. Phys.}\ }\textbf {\bibinfo {volume}
  {53}},\ \bibinfo {pages} {793} (\bibinfo {year} {2005})},\ \Eprint
  {https://arxiv.org/abs/hep-th/0502050} {arXiv:hep-th/0502050} \BibitemShut
  {NoStop}%
\bibitem [{\citenamefont {Cardoso}\ \emph
  {et~al.}(2016{\natexlab{a}})\citenamefont {Cardoso}, \citenamefont
  {Franzin},\ and\ \citenamefont {Pani}}]{Cardoso:2016}%
  \BibitemOpen
  \bibfield  {author} {\bibinfo {author} {\bibfnamefont {V.}~\bibnamefont
  {Cardoso}}, \bibinfo {author} {\bibfnamefont {E.}~\bibnamefont {Franzin}},\
  and\ \bibinfo {author} {\bibfnamefont {P.}~\bibnamefont {Pani}},\ }\href
  {https://doi.org/10.1103/PhysRevLett.116.171101} {\bibfield  {journal}
  {\bibinfo  {journal} {Phys. Rev. Lett.}\ }\textbf {\bibinfo {volume} {116}},\
  \bibinfo {pages} {171101} (\bibinfo {year} {2016}{\natexlab{a}})}\BibitemShut
  {NoStop}%
\bibitem [{\citenamefont {Cardoso}\ and\ \citenamefont
  {Pani}(2017)}]{Cardoso:2017}%
  \BibitemOpen
  \bibfield  {author} {\bibinfo {author} {\bibfnamefont {V.}~\bibnamefont
  {Cardoso}}\ and\ \bibinfo {author} {\bibfnamefont {P.}~\bibnamefont {Pani}},\
  }\href {https://doi.org/10.1038/s41550-017-0225-y} {\bibfield  {journal}
  {\bibinfo  {journal} {Nature Astron.}\ }\textbf {\bibinfo {volume} {1}},\
  \bibinfo {pages} {586} (\bibinfo {year} {2017})},\ \Eprint
  {https://arxiv.org/abs/1709.01525} {arXiv:1709.01525 [gr-qc]} \BibitemShut
  {NoStop}%
\bibitem [{\citenamefont {Cardoso}\ and\ \citenamefont
  {Pani}(2019)}]{Cardoso:2019}%
  \BibitemOpen
  \bibfield  {author} {\bibinfo {author} {\bibfnamefont {V.}~\bibnamefont
  {Cardoso}}\ and\ \bibinfo {author} {\bibfnamefont {P.}~\bibnamefont {Pani}},\
  }\href {https://doi.org/10.1007/s41114-019-0020-4} {\bibfield  {journal}
  {\bibinfo  {journal} {Living Rev. Rel.}\ }\textbf {\bibinfo {volume} {22}},\
  \bibinfo {pages} {4} (\bibinfo {year} {2019})},\ \Eprint
  {https://arxiv.org/abs/1904.05363} {arXiv:1904.05363 [gr-qc]} \BibitemShut
  {NoStop}%
\bibitem [{\citenamefont {Mazur}\ and\ \citenamefont
  {Mottola}(2001)}]{Mazur:2001}%
  \BibitemOpen
  \bibfield  {author} {\bibinfo {author} {\bibfnamefont {P.~O.}\ \bibnamefont
  {Mazur}}\ and\ \bibinfo {author} {\bibfnamefont {E.}~\bibnamefont
  {Mottola}},\ }\href {https://doi.org/10.48550/ARXIV.GR-QC/0109035} {\bibinfo
  {title} {arxiv:gr-qc/0109035}} (\bibinfo {year} {2001})\BibitemShut {NoStop}%
\bibitem [{\citenamefont {Almheiri}\ \emph {et~al.}(2013)\citenamefont
  {Almheiri}, \citenamefont {Marolf}, \citenamefont {Polchinski},\ and\
  \citenamefont {Sully}}]{Almheiri:2013}%
  \BibitemOpen
  \bibfield  {author} {\bibinfo {author} {\bibfnamefont {A.}~\bibnamefont
  {Almheiri}}, \bibinfo {author} {\bibfnamefont {D.}~\bibnamefont {Marolf}},
  \bibinfo {author} {\bibfnamefont {J.}~\bibnamefont {Polchinski}},\ and\
  \bibinfo {author} {\bibfnamefont {J.}~\bibnamefont {Sully}},\ }\bibfield
  {journal} {\bibinfo  {journal} {Journal of High Energy Physics}\ }\textbf
  {\bibinfo {volume} {2013}},\ \href {https://doi.org/10.1007/jhep02(2013)062}
  {10.1007/jhep02(2013)062} (\bibinfo {year} {2013})\BibitemShut {NoStop}%
\bibitem [{\citenamefont {Lemos}\ and\ \citenamefont
  {Zaslavskii}(2008)}]{Lemos:2008}%
  \BibitemOpen
  \bibfield  {author} {\bibinfo {author} {\bibfnamefont {J.~P.~S.}\
  \bibnamefont {Lemos}}\ and\ \bibinfo {author} {\bibfnamefont {O.~B.}\
  \bibnamefont {Zaslavskii}},\ }\href
  {https://doi.org/10.1103/PhysRevD.78.024040} {\bibfield  {journal} {\bibinfo
  {journal} {Phys. Rev. D}\ }\textbf {\bibinfo {volume} {78}},\ \bibinfo
  {pages} {024040} (\bibinfo {year} {2008})},\ \Eprint
  {https://arxiv.org/abs/0806.0845} {arXiv:0806.0845 [gr-qc]} \BibitemShut
  {NoStop}%
\bibitem [{\citenamefont {Pani}\ \emph {et~al.}(2008)\citenamefont {Pani},
  \citenamefont {Cardoso}, \citenamefont {Cadoni},\ and\ \citenamefont
  {Cavaglia}}]{Pani:2008}%
  \BibitemOpen
  \bibfield  {author} {\bibinfo {author} {\bibfnamefont {P.}~\bibnamefont
  {Pani}}, \bibinfo {author} {\bibfnamefont {V.}~\bibnamefont {Cardoso}},
  \bibinfo {author} {\bibfnamefont {M.}~\bibnamefont {Cadoni}},\ and\ \bibinfo
  {author} {\bibfnamefont {M.}~\bibnamefont {Cavaglia}},\ }\href
  {https://doi.org/10.22323/1.075.0027} {\bibfield  {journal} {\bibinfo
  {journal} {PoS}\ }\textbf {\bibinfo {volume} {BHGRS}},\ \bibinfo {pages}
  {027} (\bibinfo {year} {2008})},\ \Eprint {https://arxiv.org/abs/0901.0850}
  {arXiv:0901.0850 [gr-qc]} \BibitemShut {NoStop}%
\bibitem [{\citenamefont {Konoplya}\ and\ \citenamefont
  {Zhidenko}(2016)}]{Konoplya:2016}%
  \BibitemOpen
  \bibfield  {author} {\bibinfo {author} {\bibfnamefont {R.~A.}\ \bibnamefont
  {Konoplya}}\ and\ \bibinfo {author} {\bibfnamefont {A.}~\bibnamefont
  {Zhidenko}},\ }\href {https://doi.org/10.1088/1475-7516/2016/12/043}
  {\bibfield  {journal} {\bibinfo  {journal} {JCAP}\ }\textbf {\bibinfo
  {volume} {12}},\ \bibinfo {pages} {043}},\ \Eprint
  {https://arxiv.org/abs/1606.00517} {arXiv:1606.00517 [gr-qc]} \BibitemShut
  {NoStop}%
\bibitem [{\citenamefont {Carballo-Rubio}\ \emph {et~al.}(2018)\citenamefont
  {Carballo-Rubio}, \citenamefont {Di~Filippo}, \citenamefont {Liberati},\ and\
  \citenamefont {Visser}}]{Rubio:2018}%
  \BibitemOpen
  \bibfield  {author} {\bibinfo {author} {\bibfnamefont {R.}~\bibnamefont
  {Carballo-Rubio}}, \bibinfo {author} {\bibfnamefont {F.}~\bibnamefont
  {Di~Filippo}}, \bibinfo {author} {\bibfnamefont {S.}~\bibnamefont
  {Liberati}},\ and\ \bibinfo {author} {\bibfnamefont {M.}~\bibnamefont
  {Visser}},\ }\href {https://doi.org/10.1103/PhysRevD.98.124009} {\bibfield
  {journal} {\bibinfo  {journal} {Phys. Rev. D}\ }\textbf {\bibinfo {volume}
  {98}},\ \bibinfo {pages} {124009} (\bibinfo {year} {2018})}\BibitemShut
  {NoStop}%
\bibitem [{\citenamefont {Abedi}\ \emph {et~al.}(2017)\citenamefont {Abedi},
  \citenamefont {Dykaar},\ and\ \citenamefont {Afshordi}}]{Abedi:2016}%
  \BibitemOpen
  \bibfield  {author} {\bibinfo {author} {\bibfnamefont {J.}~\bibnamefont
  {Abedi}}, \bibinfo {author} {\bibfnamefont {H.}~\bibnamefont {Dykaar}},\ and\
  \bibinfo {author} {\bibfnamefont {N.}~\bibnamefont {Afshordi}},\ }\href
  {https://doi.org/10.1103/PhysRevD.96.082004} {\bibfield  {journal} {\bibinfo
  {journal} {Phys. Rev. D}\ }\textbf {\bibinfo {volume} {96}},\ \bibinfo
  {pages} {082004} (\bibinfo {year} {2017})},\ \Eprint
  {https://arxiv.org/abs/1612.00266} {arXiv:1612.00266 [gr-qc]} \BibitemShut
  {NoStop}%
\bibitem [{\citenamefont {Mark}\ \emph {et~al.}(2017)\citenamefont {Mark},
  \citenamefont {Zimmerman}, \citenamefont {Du},\ and\ \citenamefont
  {Chen}}]{Mark:2017}%
  \BibitemOpen
  \bibfield  {author} {\bibinfo {author} {\bibfnamefont {Z.}~\bibnamefont
  {Mark}}, \bibinfo {author} {\bibfnamefont {A.}~\bibnamefont {Zimmerman}},
  \bibinfo {author} {\bibfnamefont {S.~M.}\ \bibnamefont {Du}},\ and\ \bibinfo
  {author} {\bibfnamefont {Y.}~\bibnamefont {Chen}},\ }\href
  {https://doi.org/10.1103/PhysRevD.96.084002} {\bibfield  {journal} {\bibinfo
  {journal} {Phys. Rev. D}\ }\textbf {\bibinfo {volume} {96}},\ \bibinfo
  {pages} {084002} (\bibinfo {year} {2017})},\ \Eprint
  {https://arxiv.org/abs/1706.06155} {arXiv:1706.06155 [gr-qc]} \BibitemShut
  {NoStop}%
\bibitem [{\citenamefont {Sennett}\ \emph {et~al.}(2017)\citenamefont
  {Sennett}, \citenamefont {Hinderer}, \citenamefont {Steinhoff}, \citenamefont
  {Buonanno},\ and\ \citenamefont {Ossokine}}]{Sennett:2017}%
  \BibitemOpen
  \bibfield  {author} {\bibinfo {author} {\bibfnamefont {N.}~\bibnamefont
  {Sennett}}, \bibinfo {author} {\bibfnamefont {T.}~\bibnamefont {Hinderer}},
  \bibinfo {author} {\bibfnamefont {J.}~\bibnamefont {Steinhoff}}, \bibinfo
  {author} {\bibfnamefont {A.}~\bibnamefont {Buonanno}},\ and\ \bibinfo
  {author} {\bibfnamefont {S.}~\bibnamefont {Ossokine}},\ }\href
  {https://doi.org/10.1103/PhysRevD.96.024002} {\bibfield  {journal} {\bibinfo
  {journal} {Phys. Rev. D}\ }\textbf {\bibinfo {volume} {96}},\ \bibinfo
  {pages} {024002} (\bibinfo {year} {2017})},\ \Eprint
  {https://arxiv.org/abs/1704.08651} {arXiv:1704.08651 [gr-qc]} \BibitemShut
  {NoStop}%
\bibitem [{\citenamefont {Oshita}\ and\ \citenamefont
  {Afshordi}(2019)}]{Oshita:2018}%
  \BibitemOpen
  \bibfield  {author} {\bibinfo {author} {\bibfnamefont {N.}~\bibnamefont
  {Oshita}}\ and\ \bibinfo {author} {\bibfnamefont {N.}~\bibnamefont
  {Afshordi}},\ }\href {https://doi.org/10.1103/PhysRevD.99.044002} {\bibfield
  {journal} {\bibinfo  {journal} {Phys. Rev. D}\ }\textbf {\bibinfo {volume}
  {99}},\ \bibinfo {pages} {044002} (\bibinfo {year} {2019})},\ \Eprint
  {https://arxiv.org/abs/1807.10287} {arXiv:1807.10287 [gr-qc]} \BibitemShut
  {NoStop}%
\bibitem [{\citenamefont {Bueno}\ \emph {et~al.}(2018)\citenamefont {Bueno},
  \citenamefont {Cano}, \citenamefont {Goelen}, \citenamefont {Hertog},\ and\
  \citenamefont {Vercnocke}}]{Bueno:2017}%
  \BibitemOpen
  \bibfield  {author} {\bibinfo {author} {\bibfnamefont {P.}~\bibnamefont
  {Bueno}}, \bibinfo {author} {\bibfnamefont {P.~A.}\ \bibnamefont {Cano}},
  \bibinfo {author} {\bibfnamefont {F.}~\bibnamefont {Goelen}}, \bibinfo
  {author} {\bibfnamefont {T.}~\bibnamefont {Hertog}},\ and\ \bibinfo {author}
  {\bibfnamefont {B.}~\bibnamefont {Vercnocke}},\ }\href
  {https://doi.org/10.1103/PhysRevD.97.024040} {\bibfield  {journal} {\bibinfo
  {journal} {Phys. Rev. D}\ }\textbf {\bibinfo {volume} {97}},\ \bibinfo
  {pages} {024040} (\bibinfo {year} {2018})},\ \Eprint
  {https://arxiv.org/abs/1711.00391} {arXiv:1711.00391 [gr-qc]} \BibitemShut
  {NoStop}%
\bibitem [{\citenamefont {Cardoso}\ \emph
  {et~al.}(2016{\natexlab{b}})\citenamefont {Cardoso}, \citenamefont {Hopper},
  \citenamefont {Macedo}, \citenamefont {Palenzuela},\ and\ \citenamefont
  {Pani}}]{Cardoso_ECO:2016}%
  \BibitemOpen
  \bibfield  {author} {\bibinfo {author} {\bibfnamefont {V.}~\bibnamefont
  {Cardoso}}, \bibinfo {author} {\bibfnamefont {S.}~\bibnamefont {Hopper}},
  \bibinfo {author} {\bibfnamefont {C.~F.~B.}\ \bibnamefont {Macedo}}, \bibinfo
  {author} {\bibfnamefont {C.}~\bibnamefont {Palenzuela}},\ and\ \bibinfo
  {author} {\bibfnamefont {P.}~\bibnamefont {Pani}},\ }\href
  {https://doi.org/10.1103/PhysRevD.94.084031} {\bibfield  {journal} {\bibinfo
  {journal} {Phys. Rev. D}\ }\textbf {\bibinfo {volume} {94}},\ \bibinfo
  {pages} {084031} (\bibinfo {year} {2016}{\natexlab{b}})},\ \Eprint
  {https://arxiv.org/abs/1608.08637} {arXiv:1608.08637 [gr-qc]} \BibitemShut
  {NoStop}%
\bibitem [{\citenamefont {Dey}\ \emph {et~al.}(2020)\citenamefont {Dey},
  \citenamefont {Chakraborty},\ and\ \citenamefont {Afshordi}}]{Dey:2020lhq}%
  \BibitemOpen
  \bibfield  {author} {\bibinfo {author} {\bibfnamefont {R.}~\bibnamefont
  {Dey}}, \bibinfo {author} {\bibfnamefont {S.}~\bibnamefont {Chakraborty}},\
  and\ \bibinfo {author} {\bibfnamefont {N.}~\bibnamefont {Afshordi}},\ }\href
  {https://doi.org/10.1103/PhysRevD.101.104014} {\bibfield  {journal} {\bibinfo
   {journal} {Phys. Rev. D}\ }\textbf {\bibinfo {volume} {101}},\ \bibinfo
  {pages} {104014} (\bibinfo {year} {2020})},\ \Eprint
  {https://arxiv.org/abs/2001.01301} {arXiv:2001.01301 [gr-qc]} \BibitemShut
  {NoStop}%
\bibitem [{\citenamefont {Dey}\ \emph {et~al.}(2021)\citenamefont {Dey},
  \citenamefont {Biswas},\ and\ \citenamefont {Chakraborty}}]{Dey:2020pth}%
  \BibitemOpen
  \bibfield  {author} {\bibinfo {author} {\bibfnamefont {R.}~\bibnamefont
  {Dey}}, \bibinfo {author} {\bibfnamefont {S.}~\bibnamefont {Biswas}},\ and\
  \bibinfo {author} {\bibfnamefont {S.}~\bibnamefont {Chakraborty}},\ }\href
  {https://doi.org/10.1103/PhysRevD.103.084019} {\bibfield  {journal} {\bibinfo
   {journal} {Phys. Rev. D}\ }\textbf {\bibinfo {volume} {103}},\ \bibinfo
  {pages} {084019} (\bibinfo {year} {2021})},\ \Eprint
  {https://arxiv.org/abs/2010.07966} {arXiv:2010.07966 [gr-qc]} \BibitemShut
  {NoStop}%
\bibitem [{\citenamefont {Morris}\ \emph {et~al.}(1988)\citenamefont {Morris},
  \citenamefont {Thorne},\ and\ \citenamefont {Yurtsever}}]{Morris:1988}%
  \BibitemOpen
  \bibfield  {author} {\bibinfo {author} {\bibfnamefont {M.~S.}\ \bibnamefont
  {Morris}}, \bibinfo {author} {\bibfnamefont {K.~S.}\ \bibnamefont {Thorne}},\
  and\ \bibinfo {author} {\bibfnamefont {U.}~\bibnamefont {Yurtsever}},\ }\href
  {https://doi.org/10.1103/PhysRevLett.61.1446} {\bibfield  {journal} {\bibinfo
   {journal} {Phys. Rev. Lett.}\ }\textbf {\bibinfo {volume} {61}},\ \bibinfo
  {pages} {1446} (\bibinfo {year} {1988})}\BibitemShut {NoStop}%
\bibitem [{\citenamefont {Bronnikov}\ and\ \citenamefont
  {Kim}(2003)}]{Bronnikov:2002}%
  \BibitemOpen
  \bibfield  {author} {\bibinfo {author} {\bibfnamefont {K.~A.}\ \bibnamefont
  {Bronnikov}}\ and\ \bibinfo {author} {\bibfnamefont {S.-W.}\ \bibnamefont
  {Kim}},\ }\href {https://doi.org/10.1103/PhysRevD.67.064027} {\bibfield
  {journal} {\bibinfo  {journal} {Phys. Rev. D}\ }\textbf {\bibinfo {volume}
  {67}},\ \bibinfo {pages} {064027} (\bibinfo {year} {2003})},\ \Eprint
  {https://arxiv.org/abs/gr-qc/0212112} {arXiv:gr-qc/0212112} \BibitemShut
  {NoStop}%
\bibitem [{\citenamefont {Bronnikov}\ \emph {et~al.}(2003)\citenamefont
  {Bronnikov}, \citenamefont {Melnikov},\ and\ \citenamefont
  {Dehnen}}]{Bronnikov:2003}%
  \BibitemOpen
  \bibfield  {author} {\bibinfo {author} {\bibfnamefont {K.~A.}\ \bibnamefont
  {Bronnikov}}, \bibinfo {author} {\bibfnamefont {V.~N.}\ \bibnamefont
  {Melnikov}},\ and\ \bibinfo {author} {\bibfnamefont {H.}~\bibnamefont
  {Dehnen}},\ }\href {https://doi.org/10.1103/PhysRevD.68.024025} {\bibfield
  {journal} {\bibinfo  {journal} {Phys. Rev. D}\ }\textbf {\bibinfo {volume}
  {68}},\ \bibinfo {pages} {024025} (\bibinfo {year} {2003})}\BibitemShut
  {NoStop}%
\bibitem [{\citenamefont {Tsukamoto}\ \emph {et~al.}(2012)\citenamefont
  {Tsukamoto}, \citenamefont {Harada},\ and\ \citenamefont
  {Yajima}}]{Tsukamoto:2012}%
  \BibitemOpen
  \bibfield  {author} {\bibinfo {author} {\bibfnamefont {N.}~\bibnamefont
  {Tsukamoto}}, \bibinfo {author} {\bibfnamefont {T.}~\bibnamefont {Harada}},\
  and\ \bibinfo {author} {\bibfnamefont {K.}~\bibnamefont {Yajima}},\ }\href
  {https://doi.org/10.1103/PhysRevD.86.104062} {\bibfield  {journal} {\bibinfo
  {journal} {Phys. Rev. D}\ }\textbf {\bibinfo {volume} {86}},\ \bibinfo
  {pages} {104062} (\bibinfo {year} {2012})},\ \Eprint
  {https://arxiv.org/abs/1207.0047} {arXiv:1207.0047 [gr-qc]} \BibitemShut
  {NoStop}%
\bibitem [{\citenamefont {Ohgami}\ and\ \citenamefont
  {Sakai}(2015)}]{Ohgami:2015}%
  \BibitemOpen
  \bibfield  {author} {\bibinfo {author} {\bibfnamefont {T.}~\bibnamefont
  {Ohgami}}\ and\ \bibinfo {author} {\bibfnamefont {N.}~\bibnamefont {Sakai}},\
  }\href {https://doi.org/10.1103/PhysRevD.91.124020} {\bibfield  {journal}
  {\bibinfo  {journal} {Phys. Rev. D}\ }\textbf {\bibinfo {volume} {91}},\
  \bibinfo {pages} {124020} (\bibinfo {year} {2015})},\ \Eprint
  {https://arxiv.org/abs/1704.07065} {arXiv:1704.07065 [gr-qc]} \BibitemShut
  {NoStop}%
\bibitem [{\citenamefont {Shaikh}\ \emph {et~al.}(2019)\citenamefont {Shaikh},
  \citenamefont {Banerjee}, \citenamefont {Paul},\ and\ \citenamefont
  {Sarkar}}]{Shaikh:2018}%
  \BibitemOpen
  \bibfield  {author} {\bibinfo {author} {\bibfnamefont {R.}~\bibnamefont
  {Shaikh}}, \bibinfo {author} {\bibfnamefont {P.}~\bibnamefont {Banerjee}},
  \bibinfo {author} {\bibfnamefont {S.}~\bibnamefont {Paul}},\ and\ \bibinfo
  {author} {\bibfnamefont {T.}~\bibnamefont {Sarkar}},\ }\href
  {https://doi.org/10.1016/j.physletb.2018.12.030} {\bibfield  {journal}
  {\bibinfo  {journal} {Phys. Lett. B}\ }\textbf {\bibinfo {volume} {789}},\
  \bibinfo {pages} {270} (\bibinfo {year} {2019})},\ \bibinfo {note} {[Erratum:
  Phys.Lett.B 791, 422--423 (2019)]},\ \Eprint
  {https://arxiv.org/abs/1811.08245} {arXiv:1811.08245 [gr-qc]} \BibitemShut
  {NoStop}%
\bibitem [{\citenamefont {Banerjee}\ \emph {et~al.}(2021)\citenamefont
  {Banerjee}, \citenamefont {Paul}, \citenamefont {Shaikh},\ and\ \citenamefont
  {Sarkar}}]{Banerjee:2019}%
  \BibitemOpen
  \bibfield  {author} {\bibinfo {author} {\bibfnamefont {P.}~\bibnamefont
  {Banerjee}}, \bibinfo {author} {\bibfnamefont {S.}~\bibnamefont {Paul}},
  \bibinfo {author} {\bibfnamefont {R.}~\bibnamefont {Shaikh}},\ and\ \bibinfo
  {author} {\bibfnamefont {T.}~\bibnamefont {Sarkar}},\ }\href
  {https://doi.org/10.1088/1475-7516/2021/03/042} {\bibfield  {journal}
  {\bibinfo  {journal} {JCAP}\ }\textbf {\bibinfo {volume} {03}},\ \bibinfo
  {pages} {042}},\ \Eprint {https://arxiv.org/abs/1912.01184} {arXiv:1912.01184
  [astro-ph.HE]} \BibitemShut {NoStop}%
\bibitem [{\citenamefont {Dutta~Roy}\ \emph {et~al.}(2020)\citenamefont
  {Dutta~Roy}, \citenamefont {Aneesh},\ and\ \citenamefont
  {Kar}}]{DuttaRoy:2019}%
  \BibitemOpen
  \bibfield  {author} {\bibinfo {author} {\bibfnamefont {P.}~\bibnamefont
  {Dutta~Roy}}, \bibinfo {author} {\bibfnamefont {S.}~\bibnamefont {Aneesh}},\
  and\ \bibinfo {author} {\bibfnamefont {S.}~\bibnamefont {Kar}},\ }\href
  {https://doi.org/10.1140/epjc/s10052-020-8409-5} {\bibfield  {journal}
  {\bibinfo  {journal} {Eur. Phys. J. C}\ }\textbf {\bibinfo {volume} {80}},\
  \bibinfo {pages} {850} (\bibinfo {year} {2020})},\ \Eprint
  {https://arxiv.org/abs/1910.08746} {arXiv:1910.08746 [gr-qc]} \BibitemShut
  {NoStop}%
\bibitem [{\citenamefont {Bronnikov}\ and\ \citenamefont
  {Konoplya}(2020)}]{Bronnikov:2020}%
  \BibitemOpen
  \bibfield  {author} {\bibinfo {author} {\bibfnamefont {K.~A.}\ \bibnamefont
  {Bronnikov}}\ and\ \bibinfo {author} {\bibfnamefont {R.~A.}\ \bibnamefont
  {Konoplya}},\ }\href {https://doi.org/10.1103/PhysRevD.101.064004} {\bibfield
   {journal} {\bibinfo  {journal} {Phys. Rev. D}\ }\textbf {\bibinfo {volume}
  {101}},\ \bibinfo {pages} {064004} (\bibinfo {year} {2020})}\BibitemShut
  {NoStop}%
\bibitem [{\citenamefont {Franzin}\ \emph {et~al.}(2022)\citenamefont
  {Franzin}, \citenamefont {Liberati}, \citenamefont {Mazza}, \citenamefont
  {Dey},\ and\ \citenamefont {Chakraborty}}]{Franzin:2022}%
  \BibitemOpen
  \bibfield  {author} {\bibinfo {author} {\bibfnamefont {E.}~\bibnamefont
  {Franzin}}, \bibinfo {author} {\bibfnamefont {S.}~\bibnamefont {Liberati}},
  \bibinfo {author} {\bibfnamefont {J.}~\bibnamefont {Mazza}}, \bibinfo
  {author} {\bibfnamefont {R.}~\bibnamefont {Dey}},\ and\ \bibinfo {author}
  {\bibfnamefont {S.}~\bibnamefont {Chakraborty}},\ }\href
  {https://doi.org/10.48550/ARXIV.2201.01650} {\bibinfo {title}
  {arxiv:2201.01650}} (\bibinfo {year} {2022})\BibitemShut {NoStop}%
\bibitem [{\citenamefont {Kar}\ \emph {et~al.}(2015)\citenamefont {Kar},
  \citenamefont {Lahiri},\ and\ \citenamefont {SenGupta}}]{Kar:2015}%
  \BibitemOpen
  \bibfield  {author} {\bibinfo {author} {\bibfnamefont {S.}~\bibnamefont
  {Kar}}, \bibinfo {author} {\bibfnamefont {S.}~\bibnamefont {Lahiri}},\ and\
  \bibinfo {author} {\bibfnamefont {S.}~\bibnamefont {SenGupta}},\ }\href
  {https://doi.org/10.1016/j.physletb.2015.09.039} {\bibfield  {journal}
  {\bibinfo  {journal} {Phys. Lett. B}\ }\textbf {\bibinfo {volume} {750}},\
  \bibinfo {pages} {319} (\bibinfo {year} {2015})},\ \Eprint
  {https://arxiv.org/abs/1505.06831} {arXiv:1505.06831 [gr-qc]} \BibitemShut
  {NoStop}%
\bibitem [{\citenamefont {Randall}\ and\ \citenamefont {Sundrum}(1999)}]{RS1}%
  \BibitemOpen
  \bibfield  {author} {\bibinfo {author} {\bibfnamefont {L.}~\bibnamefont
  {Randall}}\ and\ \bibinfo {author} {\bibfnamefont {R.}~\bibnamefont
  {Sundrum}},\ }\href {https://doi.org/10.1103/PhysRevLett.83.3370} {\bibfield
  {journal} {\bibinfo  {journal} {Phys. Rev. Lett.}\ }\textbf {\bibinfo
  {volume} {83}},\ \bibinfo {pages} {3370} (\bibinfo {year} {1999})},\ \Eprint
  {https://arxiv.org/abs/hep-ph/9905221} {arXiv:hep-ph/9905221} \BibitemShut
  {NoStop}%
\bibitem [{\citenamefont {Biswas}\ \emph {et~al.}(2022)\citenamefont {Biswas},
  \citenamefont {Rahman},\ and\ \citenamefont {Chakraborty}}]{Biswas:2022}%
  \BibitemOpen
  \bibfield  {author} {\bibinfo {author} {\bibfnamefont {S.}~\bibnamefont
  {Biswas}}, \bibinfo {author} {\bibfnamefont {M.}~\bibnamefont {Rahman}},\
  and\ \bibinfo {author} {\bibfnamefont {S.}~\bibnamefont {Chakraborty}},\
  }\href {https://doi.org/10.48550/ARXIV.2205.14743} {\bibinfo {title}
  {arxiv:2205.14743}} (\bibinfo {year} {2022})\BibitemShut {NoStop}%
\bibitem [{\citenamefont {Banerjee}\ \emph
  {et~al.}(2020{\natexlab{b}})\citenamefont {Banerjee}, \citenamefont
  {Chakraborty},\ and\ \citenamefont {SenGupta}}]{Banerjee:2020}%
  \BibitemOpen
  \bibfield  {author} {\bibinfo {author} {\bibfnamefont {I.}~\bibnamefont
  {Banerjee}}, \bibinfo {author} {\bibfnamefont {S.}~\bibnamefont
  {Chakraborty}},\ and\ \bibinfo {author} {\bibfnamefont {S.}~\bibnamefont
  {SenGupta}},\ }\href {https://doi.org/10.1103/PhysRevD.101.041301} {\bibfield
   {journal} {\bibinfo  {journal} {Phys. Rev. D}\ }\textbf {\bibinfo {volume}
  {101}},\ \bibinfo {pages} {041301} (\bibinfo {year} {2020}{\natexlab{b}})},\
  \Eprint {https://arxiv.org/abs/1909.09385} {arXiv:1909.09385 [gr-qc]}
  \BibitemShut {NoStop}%
\bibitem [{\citenamefont {Boersma}\ \emph {et~al.}(2020)\citenamefont
  {Boersma}, \citenamefont {Nichols},\ and\ \citenamefont
  {Schmidt}}]{Boersma:2020}%
  \BibitemOpen
  \bibfield  {author} {\bibinfo {author} {\bibfnamefont {O.~M.}\ \bibnamefont
  {Boersma}}, \bibinfo {author} {\bibfnamefont {D.~A.}\ \bibnamefont
  {Nichols}},\ and\ \bibinfo {author} {\bibfnamefont {P.}~\bibnamefont
  {Schmidt}},\ }\href {https://doi.org/10.1103/PhysRevD.101.083026} {\bibfield
  {journal} {\bibinfo  {journal} {Phys. Rev. D}\ }\textbf {\bibinfo {volume}
  {101}},\ \bibinfo {pages} {083026} (\bibinfo {year} {2020})}\BibitemShut
  {NoStop}%
\bibitem [{\citenamefont {Braginsky}\ and\ \citenamefont
  {Grishchuk}(1985)}]{Braginsky:1985}%
  \BibitemOpen
  \bibfield  {author} {\bibinfo {author} {\bibfnamefont {V.~B.}\ \bibnamefont
  {Braginsky}}\ and\ \bibinfo {author} {\bibfnamefont {L.~P.}\ \bibnamefont
  {Grishchuk}},\ }\href@noop {} {\bibfield  {journal} {\bibinfo  {journal}
  {Sov. Phys. JETP}\ }\textbf {\bibinfo {volume} {62}},\ \bibinfo {pages} {427}
  (\bibinfo {year} {1985})}\BibitemShut {NoStop}%
\bibitem [{\citenamefont {Favata}(2010)}]{Favata:2010}%
  \BibitemOpen
  \bibfield  {author} {\bibinfo {author} {\bibfnamefont {M.}~\bibnamefont
  {Favata}},\ }\href {https://doi.org/10.1088/0264-9381/27/8/084036} {\bibfield
   {journal} {\bibinfo  {journal} {Class. Qtm. Grav.}\ }\textbf {\bibinfo
  {volume} {27}},\ \bibinfo {pages} {084036} (\bibinfo {year}
  {2010})}\BibitemShut {NoStop}%
\bibitem [{\citenamefont {H\"ubner}\ \emph {et~al.}(2021)\citenamefont
  {H\"ubner}, \citenamefont {Lasky},\ and\ \citenamefont
  {Thrane}}]{Hubner:2021}%
  \BibitemOpen
  \bibfield  {author} {\bibinfo {author} {\bibfnamefont {M.}~\bibnamefont
  {H\"ubner}}, \bibinfo {author} {\bibfnamefont {P.}~\bibnamefont {Lasky}},\
  and\ \bibinfo {author} {\bibfnamefont {E.}~\bibnamefont {Thrane}},\ }\href
  {https://doi.org/10.1103/PhysRevD.104.023004} {\bibfield  {journal} {\bibinfo
   {journal} {Phys. Rev. D}\ }\textbf {\bibinfo {volume} {104}},\ \bibinfo
  {pages} {023004} (\bibinfo {year} {2021})}\BibitemShut {NoStop}%
\bibitem [{\citenamefont {Lasky}\ \emph {et~al.}(2016)\citenamefont {Lasky},
  \citenamefont {Thrane}, \citenamefont {Levin}, \citenamefont {Blackman},\
  and\ \citenamefont {Chen}}]{Lasky:2016}%
  \BibitemOpen
  \bibfield  {author} {\bibinfo {author} {\bibfnamefont {P.~D.}\ \bibnamefont
  {Lasky}}, \bibinfo {author} {\bibfnamefont {E.}~\bibnamefont {Thrane}},
  \bibinfo {author} {\bibfnamefont {Y.}~\bibnamefont {Levin}}, \bibinfo
  {author} {\bibfnamefont {J.}~\bibnamefont {Blackman}},\ and\ \bibinfo
  {author} {\bibfnamefont {Y.}~\bibnamefont {Chen}},\ }\href
  {https://doi.org/10.1103/PhysRevLett.117.061102} {\bibfield  {journal}
  {\bibinfo  {journal} {Phys. Rev. Lett.}\ }\textbf {\bibinfo {volume} {117}},\
  \bibinfo {pages} {061102} (\bibinfo {year} {2016})}\BibitemShut {NoStop}%
\bibitem [{\citenamefont {{Zel'dovich}}\ and\ \citenamefont
  {{Polnarev}}(1974)}]{Zeldovich:1974}%
  \BibitemOpen
  \bibfield  {author} {\bibinfo {author} {\bibfnamefont {Y.~B.}\ \bibnamefont
  {{Zel'dovich}}}\ and\ \bibinfo {author} {\bibfnamefont {A.~G.}\ \bibnamefont
  {{Polnarev}}},\ }\href@noop {} {\bibfield  {journal} {\bibinfo  {journal}
  {Sov. Astron}\ }\textbf {\bibinfo {volume} {18}},\ \bibinfo {pages} {17}
  (\bibinfo {year} {1974})}\BibitemShut {NoStop}%
\bibitem [{\citenamefont {Kovacs}\ and\ \citenamefont
  {Thorne}(1978)}]{Kovacs:1978}%
  \BibitemOpen
  \bibfield  {author} {\bibinfo {author} {\bibfnamefont {S.~J.}\ \bibnamefont
  {Kovacs}}\ and\ \bibinfo {author} {\bibfnamefont {K.~S.}\ \bibnamefont
  {Thorne}},\ }\href {https://doi.org/10.1086/156350} {\bibfield  {journal}
  {\bibinfo  {journal} {Astrophys. J.}\ }\textbf {\bibinfo {volume} {224}},\
  \bibinfo {pages} {62} (\bibinfo {year} {1978})}\BibitemShut {NoStop}%
\bibitem [{\citenamefont {Christodoulou}(1991)}]{Christodoulou:1991}%
  \BibitemOpen
  \bibfield  {author} {\bibinfo {author} {\bibfnamefont {D.}~\bibnamefont
  {Christodoulou}},\ }\href {https://doi.org/10.1103/PhysRevLett.67.1486}
  {\bibfield  {journal} {\bibinfo  {journal} {Phys. Rev. Lett.}\ }\textbf
  {\bibinfo {volume} {67}},\ \bibinfo {pages} {1486} (\bibinfo {year}
  {1991})}\BibitemShut {NoStop}%
\bibitem [{\citenamefont {Bieri}\ and\ \citenamefont
  {Garfinkle}(2013)}]{Bieri:2013}%
  \BibitemOpen
  \bibfield  {author} {\bibinfo {author} {\bibfnamefont {L.}~\bibnamefont
  {Bieri}}\ and\ \bibinfo {author} {\bibfnamefont {D.}~\bibnamefont
  {Garfinkle}},\ }\href {https://doi.org/10.1088/0264-9381/30/19/195009}
  {\bibfield  {journal} {\bibinfo  {journal} {Class. Qtm. Grav.}\ }\textbf
  {\bibinfo {volume} {30}},\ \bibinfo {pages} {195009} (\bibinfo {year}
  {2013})}\BibitemShut {NoStop}%
\bibitem [{\citenamefont {Winicour}(2014)}]{Winicour:2014}%
  \BibitemOpen
  \bibfield  {author} {\bibinfo {author} {\bibfnamefont {J.}~\bibnamefont
  {Winicour}},\ }\href {https://doi.org/10.1088/0264-9381/31/20/205003}
  {\bibfield  {journal} {\bibinfo  {journal} {Class. Quant. Grav.}\ }\textbf
  {\bibinfo {volume} {31}},\ \bibinfo {pages} {205003} (\bibinfo {year}
  {2014})},\ \Eprint {https://arxiv.org/abs/1407.0259} {arXiv:1407.0259
  [gr-qc]} \BibitemShut {NoStop}%
\bibitem [{\citenamefont {Pate}\ \emph {et~al.}(2017)\citenamefont {Pate},
  \citenamefont {Raclariu},\ and\ \citenamefont {Strominger}}]{Pate:2017}%
  \BibitemOpen
  \bibfield  {author} {\bibinfo {author} {\bibfnamefont {M.}~\bibnamefont
  {Pate}}, \bibinfo {author} {\bibfnamefont {A.-M.}\ \bibnamefont {Raclariu}},\
  and\ \bibinfo {author} {\bibfnamefont {A.}~\bibnamefont {Strominger}},\
  }\href {https://doi.org/10.1103/PhysRevLett.119.261602} {\bibfield  {journal}
  {\bibinfo  {journal} {Phys. Rev. Lett.}\ }\textbf {\bibinfo {volume} {119}},\
  \bibinfo {pages} {261602} (\bibinfo {year} {2017})}\BibitemShut {NoStop}%
\bibitem [{\citenamefont {Jokela}\ \emph {et~al.}(2019)\citenamefont {Jokela},
  \citenamefont {Kajantie},\ and\ \citenamefont {Sarkkinen}}]{Jokela:2019}%
  \BibitemOpen
  \bibfield  {author} {\bibinfo {author} {\bibfnamefont {N.}~\bibnamefont
  {Jokela}}, \bibinfo {author} {\bibfnamefont {K.}~\bibnamefont {Kajantie}},\
  and\ \bibinfo {author} {\bibfnamefont {M.}~\bibnamefont {Sarkkinen}},\ }\href
  {https://doi.org/10.1103/PhysRevD.99.116003} {\bibfield  {journal} {\bibinfo
  {journal} {Phys. Rev. D}\ }\textbf {\bibinfo {volume} {99}},\ \bibinfo
  {pages} {116003} (\bibinfo {year} {2019})}\BibitemShut {NoStop}%
\bibitem [{\citenamefont {Hollands}\ \emph {et~al.}(2017)\citenamefont
  {Hollands}, \citenamefont {Ishibashi},\ and\ \citenamefont
  {Wald}}]{Hollands:2017}%
  \BibitemOpen
  \bibfield  {author} {\bibinfo {author} {\bibfnamefont {S.}~\bibnamefont
  {Hollands}}, \bibinfo {author} {\bibfnamefont {A.}~\bibnamefont
  {Ishibashi}},\ and\ \bibinfo {author} {\bibfnamefont {R.~M.}\ \bibnamefont
  {Wald}},\ }\href {https://doi.org/10.1088/1361-6382/aa777a} {\bibfield
  {journal} {\bibinfo  {journal} {Classical and Quantum Gravity}\ }\textbf
  {\bibinfo {volume} {34}},\ \bibinfo {pages} {155005} (\bibinfo {year}
  {2017})}\BibitemShut {NoStop}%
\bibitem [{\citenamefont {Satishchandran}\ and\ \citenamefont
  {Wald}(2018)}]{Wald:2018}%
  \BibitemOpen
  \bibfield  {author} {\bibinfo {author} {\bibfnamefont {G.}~\bibnamefont
  {Satishchandran}}\ and\ \bibinfo {author} {\bibfnamefont {R.~M.}\
  \bibnamefont {Wald}},\ }\href {https://doi.org/10.1103/PhysRevD.97.024036}
  {\bibfield  {journal} {\bibinfo  {journal} {Phys. Rev. D}\ }\textbf {\bibinfo
  {volume} {97}},\ \bibinfo {pages} {024036} (\bibinfo {year}
  {2018})}\BibitemShut {NoStop}%
\bibitem [{\citenamefont {Ferko}\ \emph {et~al.}(2022)\citenamefont {Ferko},
  \citenamefont {Satishchandran},\ and\ \citenamefont {Sethi}}]{Ferko:2021}%
  \BibitemOpen
  \bibfield  {author} {\bibinfo {author} {\bibfnamefont {C.}~\bibnamefont
  {Ferko}}, \bibinfo {author} {\bibfnamefont {G.}~\bibnamefont
  {Satishchandran}},\ and\ \bibinfo {author} {\bibfnamefont {S.}~\bibnamefont
  {Sethi}},\ }\href {https://doi.org/10.1103/PhysRevD.105.024072} {\bibfield
  {journal} {\bibinfo  {journal} {Phys. Rev. D}\ }\textbf {\bibinfo {volume}
  {105}},\ \bibinfo {pages} {024072} (\bibinfo {year} {2022})},\ \Eprint
  {https://arxiv.org/abs/2109.11599} {arXiv:2109.11599 [gr-qc]} \BibitemShut
  {NoStop}%
\bibitem [{\citenamefont {Du}\ and\ \citenamefont {Nishizawa}(2016)}]{Du:2016}%
  \BibitemOpen
  \bibfield  {author} {\bibinfo {author} {\bibfnamefont {S.~M.}\ \bibnamefont
  {Du}}\ and\ \bibinfo {author} {\bibfnamefont {A.}~\bibnamefont {Nishizawa}},\
  }\href {https://doi.org/10.1103/PhysRevD.94.104063} {\bibfield  {journal}
  {\bibinfo  {journal} {Phys. Rev. D}\ }\textbf {\bibinfo {volume} {94}},\
  \bibinfo {pages} {104063} (\bibinfo {year} {2016})}\BibitemShut {NoStop}%
\bibitem [{\citenamefont {Hou}\ and\ \citenamefont {Zhu}(2021)}]{Hou:2020JHEP}%
  \BibitemOpen
  \bibfield  {author} {\bibinfo {author} {\bibfnamefont {S.}~\bibnamefont
  {Hou}}\ and\ \bibinfo {author} {\bibfnamefont {Z.-H.}\ \bibnamefont {Zhu}},\
  }\href {https://doi.org/10.1007/JHEP01(2021)083} {\bibfield  {journal}
  {\bibinfo  {journal} {JHEP}\ }\textbf {\bibinfo {volume} {01}},\ \bibinfo
  {pages} {083}},\ \Eprint {https://arxiv.org/abs/2005.01310} {arXiv:2005.01310
  [gr-qc]} \BibitemShut {NoStop}%
\bibitem [{\citenamefont {Hou}(2020)}]{Hou:2020}%
  \BibitemOpen
  \bibfield  {author} {\bibinfo {author} {\bibfnamefont {S.}~\bibnamefont
  {Hou}},\ }in\ \href@noop {} {\emph {\bibinfo {booktitle} {{9th International
  Workshop on Astronomy and Relativistic Astrophysics}}}}\ (\bibinfo {year}
  {2020})\ \Eprint {https://arxiv.org/abs/2011.02087} {arXiv:2011.02087
  [gr-qc]} \BibitemShut {NoStop}%
\bibitem [{\citenamefont {{Tahura}}\ \emph {et~al.}()\citenamefont {{Tahura}},
  \citenamefont {{Nichols}}, \citenamefont {{Saffer}}, \citenamefont
  {{Stein}},\ and\ \citenamefont {{Yagi}}}]{Tahura:2020}%
  \BibitemOpen
  \bibfield  {author} {\bibinfo {author} {\bibfnamefont {S.}~\bibnamefont
  {{Tahura}}}, \bibinfo {author} {\bibfnamefont {D.~A.}\ \bibnamefont
  {{Nichols}}}, \bibinfo {author} {\bibfnamefont {A.}~\bibnamefont {{Saffer}}},
  \bibinfo {author} {\bibfnamefont {L.~C.}\ \bibnamefont {{Stein}}},\ and\
  \bibinfo {author} {\bibfnamefont {K.}~\bibnamefont {{Yagi}}},\ }\href@noop {}
  {\ }\Eprint {https://arxiv.org/abs/2007.13799} {arXiv:2007.13799 [gr-qc]}
  \BibitemShut {NoStop}%
\bibitem [{\citenamefont {Hou}\ \emph {et~al.}(2022)\citenamefont {Hou},
  \citenamefont {Zhu},\ and\ \citenamefont {Zhu}}]{Hou:2022}%
  \BibitemOpen
  \bibfield  {author} {\bibinfo {author} {\bibfnamefont {S.}~\bibnamefont
  {Hou}}, \bibinfo {author} {\bibfnamefont {T.}~\bibnamefont {Zhu}},\ and\
  \bibinfo {author} {\bibfnamefont {Z.-H.}\ \bibnamefont {Zhu}},\ }\href
  {https://doi.org/10.1103/PhysRevD.105.024025} {\bibfield  {journal} {\bibinfo
   {journal} {Phys. Rev. D}\ }\textbf {\bibinfo {volume} {105}},\ \bibinfo
  {pages} {024025} (\bibinfo {year} {2022})}\BibitemShut {NoStop}%
\bibitem [{\citenamefont {Donnay}\ \emph {et~al.}(2018)\citenamefont {Donnay},
  \citenamefont {Giribet}, \citenamefont {González},\ and\ \citenamefont
  {Puhm}}]{Donnay:2018}%
  \BibitemOpen
  \bibfield  {author} {\bibinfo {author} {\bibfnamefont {L.}~\bibnamefont
  {Donnay}}, \bibinfo {author} {\bibfnamefont {G.}~\bibnamefont {Giribet}},
  \bibinfo {author} {\bibfnamefont {H.~A.}\ \bibnamefont {González}},\ and\
  \bibinfo {author} {\bibfnamefont {A.}~\bibnamefont {Puhm}},\ }\bibfield
  {journal} {\bibinfo  {journal} {Physical Review D}\ }\textbf {\bibinfo
  {volume} {98}},\ \href {https://doi.org/10.1103/physrevd.98.124016}
  {10.1103/physrevd.98.124016} (\bibinfo {year} {2018})\BibitemShut {NoStop}%
\bibitem [{\citenamefont {Bhattacharjee}\ \emph {et~al.}(2019)\citenamefont
  {Bhattacharjee}, \citenamefont {Kumar},\ and\ \citenamefont
  {Bhattacharyya}}]{Srijit:2019}%
  \BibitemOpen
  \bibfield  {author} {\bibinfo {author} {\bibfnamefont {S.}~\bibnamefont
  {Bhattacharjee}}, \bibinfo {author} {\bibfnamefont {S.}~\bibnamefont
  {Kumar}},\ and\ \bibinfo {author} {\bibfnamefont {A.}~\bibnamefont
  {Bhattacharyya}},\ }\href {https://doi.org/10.1103/PhysRevD.100.084010}
  {\bibfield  {journal} {\bibinfo  {journal} {Phys. Rev. D}\ }\textbf {\bibinfo
  {volume} {100}},\ \bibinfo {pages} {084010} (\bibinfo {year}
  {2019})}\BibitemShut {NoStop}%
\bibitem [{\citenamefont {Rahman}\ and\ \citenamefont
  {Wald}(2020)}]{Wald:2020}%
  \BibitemOpen
  \bibfield  {author} {\bibinfo {author} {\bibfnamefont {A.~A.}\ \bibnamefont
  {Rahman}}\ and\ \bibinfo {author} {\bibfnamefont {R.~M.}\ \bibnamefont
  {Wald}},\ }\bibfield  {journal} {\bibinfo  {journal} {Physical Review D}\
  }\textbf {\bibinfo {volume} {101}},\ \href
  {https://doi.org/10.1103/physrevd.101.124010} {10.1103/physrevd.101.124010}
  (\bibinfo {year} {2020})\BibitemShut {NoStop}%
\bibitem [{\citenamefont {Zhang}\ \emph
  {et~al.}(2017{\natexlab{a}})\citenamefont {Zhang}, \citenamefont {Duval},
  \citenamefont {Gibbons},\ and\ \citenamefont {Horvathy}}]{Zhang:2017soft}%
  \BibitemOpen
  \bibfield  {author} {\bibinfo {author} {\bibfnamefont {P.-M.}\ \bibnamefont
  {Zhang}}, \bibinfo {author} {\bibfnamefont {C.}~\bibnamefont {Duval}},
  \bibinfo {author} {\bibfnamefont {G.~W.}\ \bibnamefont {Gibbons}},\ and\
  \bibinfo {author} {\bibfnamefont {P.~A.}\ \bibnamefont {Horvathy}},\ }\href
  {https://doi.org/10.1103/PhysRevD.96.064013} {\bibfield  {journal} {\bibinfo
  {journal} {Phys. Rev. D}\ }\textbf {\bibinfo {volume} {96}},\ \bibinfo
  {pages} {064013} (\bibinfo {year} {2017}{\natexlab{a}})}\BibitemShut
  {NoStop}%
\bibitem [{\citenamefont {M\"adler}\ and\ \citenamefont
  {Winicour}(2016)}]{Madler:2016}%
  \BibitemOpen
  \bibfield  {author} {\bibinfo {author} {\bibfnamefont {T.}~\bibnamefont
  {M\"adler}}\ and\ \bibinfo {author} {\bibfnamefont {J.}~\bibnamefont
  {Winicour}},\ }\href {https://doi.org/10.4249/scholarpedia.33528} {\bibfield
  {journal} {\bibinfo  {journal} {Scholarpedia}\ }\textbf {\bibinfo {volume}
  {11}},\ \bibinfo {pages} {33528} (\bibinfo {year} {2016})},\ \Eprint
  {https://arxiv.org/abs/1609.01731} {arXiv:1609.01731 [gr-qc]} \BibitemShut
  {NoStop}%
\bibitem [{\citenamefont {Kanno}\ and\ \citenamefont
  {Soda}(2002)}]{Kanno:2002iaa}%
  \BibitemOpen
  \bibfield  {author} {\bibinfo {author} {\bibfnamefont {S.}~\bibnamefont
  {Kanno}}\ and\ \bibinfo {author} {\bibfnamefont {J.}~\bibnamefont {Soda}},\
  }\href {https://doi.org/10.1103/PhysRevD.66.043526} {\bibfield  {journal}
  {\bibinfo  {journal} {Phys. Rev. D}\ }\textbf {\bibinfo {volume} {66}},\
  \bibinfo {pages} {043526} (\bibinfo {year} {2002})},\ \Eprint
  {https://arxiv.org/abs/hep-th/0205188} {arXiv:hep-th/0205188} \BibitemShut
  {NoStop}%
\bibitem [{\citenamefont {Antoniou}\ \emph {et~al.}(2020)\citenamefont
  {Antoniou}, \citenamefont {Bakopoulos}, \citenamefont {Kanti}, \citenamefont
  {Kleihaus},\ and\ \citenamefont {Kunz}}]{Antoniou:2019awm}%
  \BibitemOpen
  \bibfield  {author} {\bibinfo {author} {\bibfnamefont {G.}~\bibnamefont
  {Antoniou}}, \bibinfo {author} {\bibfnamefont {A.}~\bibnamefont
  {Bakopoulos}}, \bibinfo {author} {\bibfnamefont {P.}~\bibnamefont {Kanti}},
  \bibinfo {author} {\bibfnamefont {B.}~\bibnamefont {Kleihaus}},\ and\
  \bibinfo {author} {\bibfnamefont {J.}~\bibnamefont {Kunz}},\ }\href
  {https://doi.org/10.1103/PhysRevD.101.024033} {\bibfield  {journal} {\bibinfo
   {journal} {Phys. Rev. D}\ }\textbf {\bibinfo {volume} {101}},\ \bibinfo
  {pages} {024033} (\bibinfo {year} {2020})},\ \Eprint
  {https://arxiv.org/abs/1904.13091} {arXiv:1904.13091 [hep-th]} \BibitemShut
  {NoStop}%
\bibitem [{\citenamefont {Faraoni}(2004)}]{Faraoni:2004}%
  \BibitemOpen
  \bibfield  {author} {\bibinfo {author} {\bibfnamefont {V.}~\bibnamefont
  {Faraoni}},\ }\href {https://doi.org/10.1007/978-1-4020-1989-0} {\emph
  {\bibinfo {title} {{Cosmology in scalar tensor gravity}}}}\ (\bibinfo
  {publisher} {Kluwer Academic Publishers},\ \bibinfo {address} {Dordecht, The
  Netherlands},\ \bibinfo {year} {2004})\BibitemShut {NoStop}%
\bibitem [{\citenamefont {Dadhich}\ \emph {et~al.}(2002)\citenamefont
  {Dadhich}, \citenamefont {Kar}, \citenamefont {Mukherjee},\ and\
  \citenamefont {Visser}}]{SK:2002}%
  \BibitemOpen
  \bibfield  {author} {\bibinfo {author} {\bibfnamefont {N.}~\bibnamefont
  {Dadhich}}, \bibinfo {author} {\bibfnamefont {S.}~\bibnamefont {Kar}},
  \bibinfo {author} {\bibfnamefont {S.}~\bibnamefont {Mukherjee}},\ and\
  \bibinfo {author} {\bibfnamefont {M.}~\bibnamefont {Visser}},\ }\href
  {https://doi.org/10.1103/PhysRevD.65.064004} {\bibfield  {journal} {\bibinfo
  {journal} {Phys. Rev. D}\ }\textbf {\bibinfo {volume} {65}},\ \bibinfo
  {pages} {064004} (\bibinfo {year} {2002})}\BibitemShut {NoStop}%
\bibitem [{\citenamefont {Zhang}\ \emph
  {et~al.}(2017{\natexlab{b}})\citenamefont {Zhang}, \citenamefont {Duval},
  \citenamefont {Gibbons},\ and\ \citenamefont {Horvathy}}]{Zhang:2017}%
  \BibitemOpen
  \bibfield  {author} {\bibinfo {author} {\bibfnamefont {P.-M.}\ \bibnamefont
  {Zhang}}, \bibinfo {author} {\bibfnamefont {C.}~\bibnamefont {Duval}},
  \bibinfo {author} {\bibfnamefont {G.~W.}\ \bibnamefont {Gibbons}},\ and\
  \bibinfo {author} {\bibfnamefont {P.~A.}\ \bibnamefont {Horvathy}},\ }\href
  {https://doi.org/10.1016/j.physletb.2017.07.050} {\bibfield  {journal}
  {\bibinfo  {journal} {Phys. Lett. B}\ }\textbf {\bibinfo {volume} {772}},\
  \bibinfo {pages} {743} (\bibinfo {year} {2017}{\natexlab{b}})}\BibitemShut
  {NoStop}%
\bibitem [{\citenamefont {Flanagan}\ \emph {et~al.}(2019)\citenamefont
  {Flanagan}, \citenamefont {Grant}, \citenamefont {Harte},\ and\ \citenamefont
  {Nichols}}]{Flanagan:2019}%
  \BibitemOpen
  \bibfield  {author} {\bibinfo {author} {\bibfnamefont {E.~E.}\ \bibnamefont
  {Flanagan}}, \bibinfo {author} {\bibfnamefont {A.~M.}\ \bibnamefont {Grant}},
  \bibinfo {author} {\bibfnamefont {A.~I.}\ \bibnamefont {Harte}},\ and\
  \bibinfo {author} {\bibfnamefont {D.~A.}\ \bibnamefont {Nichols}},\ }\href
  {https://doi.org/10.1103/PhysRevD.99.084044} {\bibfield  {journal} {\bibinfo
  {journal} {Phys. Rev. D}\ }\textbf {\bibinfo {volume} {99}},\ \bibinfo
  {pages} {084044} (\bibinfo {year} {2019})}\BibitemShut {NoStop}%
\bibitem [{\citenamefont {Chakraborty}\ and\ \citenamefont
  {Kar}(2020{\natexlab{a}})}]{Chak:2020}%
  \BibitemOpen
  \bibfield  {author} {\bibinfo {author} {\bibfnamefont {I.}~\bibnamefont
  {Chakraborty}}\ and\ \bibinfo {author} {\bibfnamefont {S.}~\bibnamefont
  {Kar}},\ }\href {https://doi.org/10.1103/PhysRevD.101.064022} {\bibfield
  {journal} {\bibinfo  {journal} {Phys. Rev. D}\ }\textbf {\bibinfo {volume}
  {101}},\ \bibinfo {pages} {064022} (\bibinfo {year}
  {2020}{\natexlab{a}})}\BibitemShut {NoStop}%
\bibitem [{\citenamefont {Chakraborty}\ and\ \citenamefont
  {Kar}(2020{\natexlab{b}})}]{Chak1:2020}%
  \BibitemOpen
  \bibfield  {author} {\bibinfo {author} {\bibfnamefont {I.}~\bibnamefont
  {Chakraborty}}\ and\ \bibinfo {author} {\bibfnamefont {S.}~\bibnamefont
  {Kar}},\ }\href {https://doi.org/10.1016/j.physletb.2020.135611} {\bibfield
  {journal} {\bibinfo  {journal} {Phys. Lett. B}\ }\textbf {\bibinfo {volume}
  {808}},\ \bibinfo {pages} {135611} (\bibinfo {year}
  {2020}{\natexlab{b}})}\BibitemShut {NoStop}%
\bibitem [{\citenamefont {Siddhant}\ \emph {et~al.}(2021)\citenamefont
  {Siddhant}, \citenamefont {Chakraborty},\ and\ \citenamefont
  {Kar}}]{Siddhant:2020}%
  \BibitemOpen
  \bibfield  {author} {\bibinfo {author} {\bibfnamefont {S.}~\bibnamefont
  {Siddhant}}, \bibinfo {author} {\bibfnamefont {I.}~\bibnamefont
  {Chakraborty}},\ and\ \bibinfo {author} {\bibfnamefont {S.}~\bibnamefont
  {Kar}},\ }\href {https://doi.org/10.1140/epjc/s10052-021-09118-4} {\bibfield
  {journal} {\bibinfo  {journal} {Eur. Phys. J. C}\ }\textbf {\bibinfo {volume}
  {81}},\ \bibinfo {pages} {350} (\bibinfo {year} {2021})}\BibitemShut
  {NoStop}%
\bibitem [{\citenamefont {Chakraborty}(2022)}]{Chak:2022}%
  \BibitemOpen
  \bibfield  {author} {\bibinfo {author} {\bibfnamefont {I.}~\bibnamefont
  {Chakraborty}},\ }\href {https://doi.org/10.1103/PhysRevD.105.024063}
  {\bibfield  {journal} {\bibinfo  {journal} {Phys. Rev. D}\ }\textbf {\bibinfo
  {volume} {105}},\ \bibinfo {pages} {024063} (\bibinfo {year}
  {2022})}\BibitemShut {NoStop}%
\bibitem [{\citenamefont {Bondi}\ \emph {et~al.}(1962)\citenamefont {Bondi},
  \citenamefont {van~der Burg},\ and\ \citenamefont {Metzner}}]{Bondi:1962}%
  \BibitemOpen
  \bibfield  {author} {\bibinfo {author} {\bibfnamefont {H.}~\bibnamefont
  {Bondi}}, \bibinfo {author} {\bibfnamefont {M.~G.~J.}\ \bibnamefont {van~der
  Burg}},\ and\ \bibinfo {author} {\bibfnamefont {A.~W.~K.}\ \bibnamefont
  {Metzner}},\ }\href {https://doi.org/10.1098/rspa.1962.0161} {\bibfield
  {journal} {\bibinfo  {journal} {Proc. Roy. Soc. Lond. A}\ }\textbf {\bibinfo
  {volume} {269}},\ \bibinfo {pages} {21} (\bibinfo {year} {1962})}\BibitemShut
  {NoStop}%
\bibitem [{\citenamefont {Hawking}\ \emph {et~al.}(2016)\citenamefont
  {Hawking}, \citenamefont {Perry},\ and\ \citenamefont
  {Strominger}}]{Hawking:2016}%
  \BibitemOpen
  \bibfield  {author} {\bibinfo {author} {\bibfnamefont {S.~W.}\ \bibnamefont
  {Hawking}}, \bibinfo {author} {\bibfnamefont {M.~J.}\ \bibnamefont {Perry}},\
  and\ \bibinfo {author} {\bibfnamefont {A.}~\bibnamefont {Strominger}},\
  }\bibfield  {journal} {\bibinfo  {journal} {Physical Review Letters}\
  }\textbf {\bibinfo {volume} {116}},\ \href
  {https://doi.org/10.1103/physrevlett.116.231301}
  {10.1103/physrevlett.116.231301} (\bibinfo {year} {2016})\BibitemShut
  {NoStop}%
\bibitem [{\citenamefont {Kolekar}\ and\ \citenamefont
  {Louko}(2017)}]{Kolekar:2017}%
  \BibitemOpen
  \bibfield  {author} {\bibinfo {author} {\bibfnamefont {S.}~\bibnamefont
  {Kolekar}}\ and\ \bibinfo {author} {\bibfnamefont {J.}~\bibnamefont
  {Louko}},\ }\href {https://doi.org/10.1103/PhysRevD.96.024054} {\bibfield
  {journal} {\bibinfo  {journal} {Phys. Rev. D}\ }\textbf {\bibinfo {volume}
  {96}},\ \bibinfo {pages} {024054} (\bibinfo {year} {2017})},\ \Eprint
  {https://arxiv.org/abs/1703.10619} {arXiv:1703.10619 [hep-th]} \BibitemShut
  {NoStop}%
\end{thebibliography}%

\end{document}